# Analytical Model for the Tidal Evolution of the Evection Resonance and the Timing of Resonance Escape

**William R. Ward, Robin M. Canup, and Raluca Rufu**

Planetary Science Directorate Southwest Research Institute; Boulder, CO 80302

Corresponding author: Robin M. Canup (robin@boulder.swri.edu)

**Key Points:**

- Capture of the Moon into the evection resonance with the Sun transfers angular momentum from the Earth-Moon to Earth's heliocentric orbit.

- Using the Mignard tidal model, we find that escape from evection occurs early with minimal angular momentum loss from the Earth-Moon.

- Processes beyond formal evection resonance are needed to reconcile a high-angular momentum giant impact with the current Earth-Moon.

**Abstract**

A high-angular momentum giant impact with the Earth can produce a Moon with a silicate isotopic composition nearly identical to that of Earth's mantle, consistent with observations of terrestrial and lunar rocks. However, such an event requires subsequent angular momentum removal for consistency with the current Earth-Moon system. The early Moon may have been captured into the evection resonance, occurring when the lunar perigee precession period equals one year. It has been proposed that after a high-angular momentum giant impact, evection removed the angular momentum excess from the Earth-Moon pair and transferred it to Earth's orbit about the Sun. However, prior N-body integrations suggest this result depends on the tidal model and chosen tidal parameters. Here we examine the Moon's encounter with evection using a complementary analytic description and the Mignard tidal model. While the Moon is in resonance the lunar longitude of perigee librates, and if tidal evolution excites the libration amplitude sufficiently, escape from resonance occurs. The angular momentum drain produced by formal evection depends on how long the resonance is maintained. We estimate that resonant escape occurs early, leading to only a small reduction (~ few to 10%) in the Earth-Moon system angular momentum. Moon formation from a high-angular momentum impact would then require other angular momentum removal mechanisms beyond standard libration in evection, as have been suggested previously.

**Plain Language Summary**

A canonical giant impact with the Earth by a Mars-sized impactor can produce the Moon and the current Earth-Moon angular momentum. However, such an impact would produce a planet and protolunar disk with very different proportions of impactor-derived material, likely leading to Earth-Moon compositional differences that are inconsistent with observed Earth-Moon isotopic similarities. Alternatively, a high-angular momentum impact could form a disk with a silicate composition similar to that of the Earth, but with a post-impact angular momentum much higher than in the current Earth-Moon system. As the early Moon tidally receded from the Earth, its perigee precession period lengthened. When this period equaled one year, the Moon may have been captured into the evection resonance with the Sun. It has been proposed that evection removed the angular momentum excess from the Earth-Moon pair, but the appropriate degree of angular momentum removal appears sensitive to tidal models. In this work, we use an analytical model to examine the Moon's evolution in evection and find that escape from formal resonance occurs early, with limited angular momentum reduction. Thus, in order for a high-angular momentum giant impact to be consistent with the current Earth-Moon system, additional mechanisms that do not involve standard resonance occupancy appears required.

# 1 Introduction

The leading theory for lunar origin proposes that the Moon formed from material ejected into circumterrestrial orbit by a Mars-sized impactor colliding obliquely with the early Earth (Cameron and Ward, 1976). The impact theory became favored primarily for its ability to account for the Moon's depletion in iron and the angular momentum of the current Earth-Moon system, $L_{\text{EM}} = 3.5 \times 10^{41} \text{ g cm}^2 \text{ s}^{-2}$, with the latter implying an Earth day of about 5 hours when the Moon formed close to the Earth.

In what is sometimes referred to as the "canonical" case, a low-velocity, oblique impact of a Mars-sized body produces an Earth-disk system with an angular momentum close to $L_{\text{EM}}$ (*e.g.*, Canup and Asphaug, 2001; Canup, 2004a, b, 2008). Disks produced by canonical impacts are derived primarily from material originating in the impactor's mantle. The isotopic composition of the impactor would have likely differed from that of the Earth (Pahlevan and Stevenson, 2007; Melosh, 2014; Kruijer and Kleine, 2017). Thus a disk derived from the impactor would nominally yield a Moon whose composition differed from that of the Earth's mantle. Instead, the Moon and the silicate Earth have essentially identical isotopic compositions across all non-volatile elements, including oxygen, chromium, titanium, silicon, and tungsten (*e.g.*, Lugmair and Shukolyukov, 1998; Wiechert et al., 2001; Touboul *et al.*, 2007; Armytage *et al.*, 2012; Zhang *et al.*, 2012; Kruijer *et al.,* 2015; Touboul *et al.,* 2015). Thus, a canonical impact appears to either require a low-probability compositional match between the impactor and Earth (*e.g.*, Kruijer and Kleine, 2017), or that the disk and the post-impact Earth mixed and compositionally equilibrated after the impact but before the Moon formed (Pahlevan and Stevenson, 2007; Lock *et al.*, 2018).

Alternatively, certain types of high angular momentum impacts can directly produce a protolunar disk whose silicate composition is essentially identical to that of the Earth's mantle (Ćuk and Stewart, 2012; Canup, 2012), accounting for nearly all Earth-Moon isotopic similarities without requiring an Earth-like impactor. Recent works on high-angular momentum impacts in general suggest that due to the high energy of such events, the collisional aftermath can consist of a hot, pressure supported planet rotating uniformly out to its corotation limit, while beyond that the structure progressively transitions to a disk with a keplerian profile (Lock and Stewart, 2017; Lock *et al.,*2018). It is argued that such structures, termed "synestias", may have intermittently existed during the planetary accretion process and would have facilitated the formation of moons with compositions similar to that of their host planet via mixing and equilibration. However, all high-angular momentum impacts leave the Earth-Moon system with a substantial angular momentum excess compared to its current value, so that relevancy to lunar origin requires a reliable mechanism(s) to subsequently reduce the system angular momentum.

Tidal interactions between the Earth and Moon conserve angular momentum, but other processes can remove angular momentum from the pair. Consider an initial lunar orbit that lies within the Earth's equatorial plane and a lunar spin axis normal to that plane. The energy, *E,* and scalar angular momentum, *L,* of the Earth-Moon system are

$$E = \frac{1}{2}Cs^2 + \frac{1}{2}C_m s_m^2 - \frac{GMm}{2a} \quad (1.1)$$

$$L = Cs + C_m s_m + m(GMa)^{1/2}(1-e^2)^{1/2} \quad (1.2)$$

where $M = 5.97 \times 10^{27}$ g and $m = 7.34 \times 10^{25}$ g are the Earth and lunar masses, $(C, s)$ and $(C_m, s_m)$ are their principal moments of inertia and spin rates, respectively, and the final terms are the energy, $E_{\text{orb}}$, and angular momentum, $L_{\text{orb}}$, of the lunar orbit, with $a$ and $e$ being its semi-major axis and eccentricity (see Table 1 for variable definitions). Over the age of the solar system, *L* has decreased due to a slowdown in the Earth's spin caused by direct solar tides. Additionally, late-veneer impacts could stochastically change the angular momentum of the Earth-Moon system (Bottke *et al.*, 2010). However, these processes are thought to induce only small changes (by a few to 10%) insufficient to reconcile a high-angular momentum impact with the current Earth-Moon.[1]

[Table 1]

---

[1] Angular momentum can also be lost if during its accretion, the Moon scatters some disk material onto orbits that escape the Earth. However, lunar accretion models suggest this leads to only a minimal reduction of order a few percent (*e.g.*, Kokubo *et al.*, 2000; Salmon and Canup, 2012).

The solar influence on the Earth-Moon system is not necessarily limited to its tides, and another angular momentum removal process involves a resonance with the Sun. As the early Moon's orbit expands due to tidal interaction with the Earth, it can be captured into the evection resonance, which occurs when the precession frequency of the Moon's perigee, $\varpi$, equals that of the Earth's solar orbit, $\Omega_\odot = 1.99 \times 10^{-7} \text{s}^{-1}$ (*e.g.,* Brouwer and Clemence, 1961; Kaula and Yoder, 1976; Touma and Wisdom, 1998). Capture into evection excites the Moon's orbital eccentricity and drains angular momentum from the Earth-Moon pair, transferring it to Earth's heliocentric orbit. Once the lunar eccentricity becomes sufficiently high, there is a phase during which the lunar orbit temporarily contracts due to the effects of lunar tides. Touma and Wisdom (1998) modeled capture of the Moon in evection for an initial terrestrial rotation period of 5 hr, assuming a lunar rotation synchronous with its mean motion and the Mignard tidal model, in which the tidal distortion forms some fixed time after the tide raising potential (see Section 4.1). In their simulations, the Moon's residence in evection is brief, with escape occurring soon after the lunar semi-major axis begins to contract, leading to only minor angular momentum modification by a few percent of $L_{EM}$ (*e.g.*, Canup, 2008).

A key development was the work of Ćuk and Stewart (2012, hereafter CS12), who argued that if the original magnitude of $L$ was substantially greater than at present, prolonged capture in evection could have reduced the Earth-Moon system angular momentum by a factor of two or more. This would make it viable for a high-angular momentum impact to have produced the Moon. CS12 utilized an ersatz tidal model intended to approximate a constant lag angle/constant-$Q$ model, and considered an initial terrestrial day of only 2 to 3 hr, which shifts the position of evection outward in orbital radius relative to the cases in Touma and Wisdom (1998). The CS12 simulations showed a protracted residence of the Moon in evection that persisted even as the Moon's orbit contracted, and that during the contraction phase, large-scale angular momentum removal comparable in magnitude to $L_{EM}$ occurred. In their simulations, the final system angular momentum (AM) when the Moon escapes from resonance depends on the relative strength of tidal dissipation in the Moon compared with that in the Earth, with the final angular momentum achieving a minimum value close to $L_{EM}$ across a relatively narrow range of this ratio. Their results thus implied that for certain tidal parameters, a final angular momentum $\approx L_{EM}$ would be the limiting post-evection system value, independent of the starting angular momentum.

Considering the importance of this issue to the origin of the Moon, and that angular momentum removal due to evection depends on the Moon's tidal evolution, it is imperative to understand the robustness of AM removal for other tidal models. Wisdom and Tian (2015) demonstrated that substantial differences in the angular momentum removed compared with CS12 occur when a full constant-$Q$ Darwin-Kaula model is applied (Kaula, 1964). They instead identified a "limit cycle" in which the system circulates around the stationary points associated with evection and appropriate AM can be lost even though the evection resonance angle is not librating, although this again appeared to require a relatively narrow range of tidal parameters (Wisdom and Tian, 2015; Tian *et al.*, 2017). Further work that included the effects of tidal heating within an eccentrically orbiting Moon on the lunar tidal dissipation properties concluded that the evection resonance proper does not remove substantial AM, but that the limit cycle can (Tian *et al.*, 2017).

In this paper, we examine the Moon's evolution in evection using the Mignard tidal model as in Touma and Wisdom (1998), but we apply it to the higher angular momentum systems considered in CS12 and consider the Moon's potentially non-synchronous rotation. All common tidal models have approximations and uncertainties. A strength of the Mignard model is its straightforward analytic form, whose tidal acceleration varies smoothly near synchronous orbit and for highly-eccentric orbits in a physically intuitive manner. For the post-giant impact, fluid-like Earth (*e.g.*, Zahnle *et al.*, 2015), it seems reasonable that the position of the terrestrial equilibrium tide would reflect a characteristic time for the tide to form in the presence of internal dissipation, as assumed in the Mignard model, rather than a characteristic fixed angle relative to the Moon's position, as assumed in the constant-$Q$ tidal model. However, it is also the case that for the current Moon, the lunar tidal $Q$ value does not display the inverse frequency dependence consistent with a time delay model, but instead varies weakly with frequency (*e.g.*, Williams and Boggs, 2014; Wisdom and Tian, 2015). In any case, the Mignard model permits a detailed examination of Earth-Moon-Sun interactions during the tidal evolution of the evection resonance to test whether the behavior first described in CS12 occurs with this model as well, and in so doing, to better understand the likelihood of large-scale modification of the Earth-Moon system angular momentum.

## 2 Evection

We assume the Moon forms interior to the evection resonance on a low eccentricity orbit and then tidally evolves outward until it reaches the resonance site, $a_{\text{res}}$, where the lunar apsidal precession rate equals the frequency of the Earth's orbit. Because the lunar precession rate is a function of the Earth's oblateness, which is in turn a function of Earth's spin rate, the resonance location depends on Earth's spin rate when the Moon formed. For an initial Earth-Moon angular momentum, $L_0$, equal to that in the current Earth-Moon system ($L_{\text{EM}}$), evection is first encountered at $a_{\text{res}} \sim 4.6R$, where $R$ is the Earth's radius (*e.g.*, Touma and Wisdom, 1998). An initial high-AM system with $L_o \sim 2L_{\text{EM}}$ leads to $a_{\text{res}} \sim 7R$ (*e.g.*, CS12).

### 2.1 Lagrange Equations

We consider Earth on a circular orbit with zero obliquity, and that the initial lunar inclination is negligible, so that the terrestrial and lunar orbits are co-planar. The disturbing function of the Sun acting on the Moon up to the second order Legendre polynomial and including only the oscillating term due to evection is (*e.g.*, Brouwer and Clemence, 1961; Frouard *et al.*, 2010)

$$\Phi_\odot = -(a\Omega_\odot)^2 \left[\frac{1}{4} + \frac{3}{8}e^2 + \frac{15}{8}e^2 \cos 2\varphi\right], \tag{2.1}$$

where $\varphi \equiv \varpi - \lambda_\odot$ is the resonance phase angle, $\varpi$ is the Moon's longitude of perigee, and $\lambda_\odot$ is the solar longitude. The secular part of the potential for the Earth's quadrupole field is (*e.g.*, Efroimsky, 2005),

$$\Phi_\oplus = -\frac{1}{2}\frac{GM}{a}\frac{J_2(R/a)^2}{(1-e^2)^{3/2}}, \tag{2.2}$$

where $J_2$ is the second order gravity coefficient. From Lagrange's equations,

$$\frac{de}{dt} = \frac{(1-e^2)^{1/2}}{na^2 e}\frac{\partial \Phi_\odot}{\partial \varpi} = \frac{15}{4}e(1-e^2)^{1/2}\Omega_\odot\left(\frac{\Omega_\odot}{n}\right)\sin 2\varphi, \tag{2.3}$$

$$\frac{d\varpi}{dt} = -\frac{(1-e^2)^{1/2}}{na^2 e}\frac{\partial}{\partial e}(\Phi_\oplus + \Phi_\odot) = \frac{3}{2}n\frac{J_2(R/a)^2}{(1-e^2)^2} + \frac{3}{4}(1-e^2)^{1/2}\Omega_\odot\left(\frac{\Omega_\odot}{n}\right)(1 + 5\cos 2\varphi), \tag{2.4}$$

where $n = \sqrt{GM/a^3}$ is the lunar mean motion. The apsidal precession rate is dominated by the Earth's quadrupole and increases with the lunar eccentricity. In the vicinity of evection, $\dot{\varpi}$ approaches $\Omega_\odot$ and the phase angle, $\varphi$, changes slowly. The potential is stationary in a reference frame rotating with the Sun, so that in the absence of tides, an integral of the motion is given by the Jacobi constant (Appendix A),

$$J = E_{orb} + \Phi_\oplus + \Phi_\odot - \Omega_\odot L_{orb} = m\left[-GM/2a + \Phi_\oplus + \Phi_\odot - \Omega_\odot\sqrt{GMa(1-e^2)}\right] \quad (2.5)$$

## 2.2 Normalized Forms

We normalize energy to $MR^2\Omega_\oplus^2$, angular momentum to $C\Omega_\oplus$, where $\Omega_\oplus \equiv (GM/R^3)^{1/2} = 1.24 \times 10^{-3} s^{-1}$ is the orbital frequency at the surface of the Earth, and the semi-major axis to Earth radii. In these units, a scaled Earth spin angular momentum of unity corresponds to rotation at approximately the stability limit. Eqns. (1.1) and (1.2) become

$$E' = \tfrac{\lambda}{2}s'^2 + \kappa\tfrac{\lambda}{2}s'^2_m - \tfrac{\mu}{2a'}, \quad (2.6)$$

$$L' = s' + \kappa s'_m + \gamma a'^{1/2}(1-e^2)^{1/2} \quad (2.7)$$

where $s' \equiv s/\Omega_\oplus$, $s'_m \equiv s_m/\Omega_\oplus$, and $a' \equiv a/R$. Here $\gamma \equiv \mu/\lambda = 0.0367$, $\mu \equiv m/M = 0.0123$ is the Moon-Earth mass ratio, and $\lambda \equiv C/MR^2 = 0.335$ is Earth's gyration constant. The quantity $\kappa \equiv C_m/C = 1.07 \times 10^{-3}$ is the ratio of maximum principal moments of inertia of the two bodies, while the final term of (2.7) is the normalized orbital angular momentum of the Moon, $L'_{orb} \equiv \gamma a'^{1/2}(1-e^2)^{1/2}$. The equations for $\dot{e}$ and $\dot{\varphi} = \dot{\varpi} - \Omega_\odot$ take the non-dimensional forms

$$\frac{de}{d\tau} = \tfrac{15}{4}\chi e(1-e^2)^{1/2}a'^{3/2}\left(\tfrac{\Omega_\odot}{\Omega_\oplus}\right)\sin 2\varphi, \quad (2.8)$$

$$\frac{d\varphi}{d\tau} = \chi\left[\frac{\Lambda^2 s'^2}{a'^{7/2}(1-e^2)^2} - 1 + \tfrac{3}{4}(1-e^2)^{1/2}a'^{3/2}\left(\tfrac{\Omega_\odot}{\Omega_\oplus}\right)(1+5\cos 2\varphi)\right]. \quad (2.9)$$

Here we set $J_2 = J_*s'^2$ to approximate the effect of the Earth's spin on its oblateness, defined $\Lambda \equiv [(3/2)J_*\Omega_\oplus/\Omega_\odot]^{1/2}$ and $\chi \equiv \Omega_\odot t_T$, and introduced a normalized time, $\tau \equiv t/t_T$, referenced to a tidal timescale, $t_T$, that will depend on the tidal model. Numerical values are $\Omega_\odot/\Omega_\oplus = 1.61 \times 10^{-4}$, $J_* = 0.315$, and $\Lambda = 54.2$.

The Jacobi constant can also be written in a non-dimensional form. The solar terms alter $e$ and $\varphi$ but do not change the energy of the Earth-Moon system, so that the Moon's semi-major axis and the spin of the Earth are constants in the absence of tides. We scale $J$ by $mR^2\Omega_\oplus^2$, and then rearrange terms that do not depend on $e$ or $\varphi$ to define $J' \equiv -(J/mR^2\Omega_\oplus^2 + 1/2a')(\Omega_\oplus/\Omega_\odot)/a'^{1/2} - (\Omega_\odot/\Omega_\oplus)a'^{3/2}/4$, which will be a constant in the absence of tides. This constant is given by

$$J' = \tfrac{1}{3}\frac{\Lambda^2 s'^2}{a'^{7/2}(1-e^2)^{3/2}} + (1-e^2)^{1/2} + \tfrac{3}{8}\tfrac{\Omega_\odot}{\Omega_\oplus}a'^{3/2}e^2(1+5\cos 2\varphi). \quad (2.10)$$

Since the equations of motion depend on the square of the eccentricity, we introduce the variable $\varepsilon = e^2$, as well as the angle, $\theta \equiv \varphi - \pi/2$, which is the libration angle relative to the positive $y$-axis in the direction of the negative $x$-axis (with the Sun positioned along the positive $x$-axis), so that $\theta = 0, \pi$ correspond to the stable points for evection (see below). Finally, we define the quantities

$$\eta \equiv \Lambda s'/a'^{7/4} \quad ; \quad \alpha \equiv \alpha_o a'^{3/2} \quad ; \quad \alpha_o \equiv (3/8)\Omega_\odot/\Omega_\oplus, \quad (2.11)$$

with $\alpha_o = 6.04 \times 10^{-5}$. The evolution equations due to evection and the related Jacobi constant simplify to

$$\dot{\varepsilon} = -20\chi\alpha\varepsilon(1-\varepsilon)^{1/2}\sin 2\theta, \quad (2.12)$$

$$\dot{\theta} = \chi\left[\eta^2/(1-\varepsilon)^2 - 1 + 2\alpha(1-\varepsilon)^{1/2}(1-5\cos 2\theta)\right], \quad (2.13)$$

$$J' = \eta^2/[3(1-\varepsilon)^{3/2}] + (1-\varepsilon)^{1/2} + \alpha\varepsilon(1-5\cos 2\theta). \quad (2.14)$$

## 2.3 Stationary States

Stationary points of the resonance occur where $\dot{\varepsilon} = \dot{\theta} = 0$. The value $\dot{\varepsilon}$ vanishes at $\theta = 0, \pi/2, \pi$, and $3\pi/2$, while $\dot{\theta} = 0$ occurs when

$$(1-\varepsilon)\left[1 - 2\alpha(1 - 5\cos 2\theta)(1-\varepsilon)^{1/2}\right]^{1/2} = \eta \ . \qquad (2.15)$$

When the resonance is fully developed, there are four stationary points at $(\varepsilon, \theta) = (\varepsilon_s, 0), (\varepsilon_s, \pi)$ and $(\varepsilon_{sx}, \pm \pi/2)$. Finally, $\varepsilon = 0$ is also a stationary point since it implies $\dot{\varepsilon} = 0$, although in this case the angle $\theta$ is degenerate.

## 2.4 Expansion to $\mathcal{O}(e^4)$

We now expand the governing expressions to $\mathcal{O}(e^4) = \mathcal{O}(\varepsilon^2)$, a reasonable approximation for $\varepsilon < 0.4$ (i.e., $e \leq 0.6$) that provides sufficient accuracy to capture the relevant behavior (e.g., Touma and Wisdom, 1998; Murray and Dermott, 1999). The variable $\alpha$ is small, of order few $\times 10^{-4}$ to $5 \times 10^{-3}$ for $3 < a' < 20$. Expanding eqn. (2.15) to lowest order in $\alpha$ gives $\varepsilon \approx 1 - \eta - \alpha(1 - 5\cos 2\theta)\eta^{3/2}$, and by further neglecting $\mathcal{O}(\alpha\varepsilon)$ terms (that will typically be smaller than $\mathcal{O}(\varepsilon^2)$ terms), we find approximate expressions for the stationary points (Appendix B),

$$\varepsilon_s = \varepsilon_* + 5\alpha \ ; \ \varepsilon_{sx} = \varepsilon_* - 5\alpha \ , \qquad (2.16)$$

whose average is

$$\varepsilon_* \approx 1 - \eta - \alpha \ , \qquad (2.17)$$

which in turn implies $1 - \eta^2 = 1 - [(1-\varepsilon_*) - \alpha]^2 \approx 2(1-\eta) - \varepsilon_*^2$. Expanding eqn. (2.14) and rearranging gives

$$J' - 1 - \eta^2/3 = (5\eta^2 - 1)\varepsilon^2/8 - (1 - \eta^2 - 2\alpha + 10\alpha\cos 2\theta)\varepsilon/2 \equiv \tilde{J}. \qquad (2.18)$$

Consistent with $\mathcal{O}(\varepsilon^2)$ accuracy, we further simplify (2.18) by setting $\eta \to 1$ in the coefficient of $\varepsilon^2$, $1 - \eta^2 \approx 2(1-\eta)$ in the $\varepsilon$ coefficient, and defining

$$\beta \equiv 1 - \eta - \alpha(1 - 5\cos 2\theta) \approx \varepsilon_* + 5\alpha\cos 2\theta \ , \qquad (2.19)$$

so that

$$\tilde{J} \approx (\varepsilon - 2\beta)\varepsilon/2 \ ; \ \varepsilon = \beta \pm \sqrt{\beta^2 + 2\tilde{J}} \ , \qquad (2.20\text{a,b})$$

where the first expression gives the Jacobi constant to $\mathcal{O}(\varepsilon^2)$, and the second gives the solutions for $\varepsilon(\theta)$ from this quadratic equation. The rates of change for the eccentricity and resonance angle that are compatible with this approximation become

$$\dot{\varepsilon} = -20\chi\alpha\varepsilon\sin 2\theta \ ; \ \dot{\theta} = 2\chi(\varepsilon - \varepsilon_* - 5\alpha\cos 2\theta) = 2\chi(\varepsilon - \beta) \ , \qquad (2.21\text{a,b})$$

where in the last expression we have dropped a term $\chi(\varepsilon_*^2 - \varepsilon(4\varepsilon_* - 3\varepsilon))$, because for an eccentricity similar to that of the stationary point, i.e., $\varepsilon \sim \varepsilon_* \pm 5\alpha$, $(\varepsilon_*^2 - \varepsilon(4\varepsilon_* - 3\varepsilon))$ is $\mathcal{O}(\alpha\varepsilon)$. We utilize eqns. (2.21a,b) to describe the effects of evection on the system evolution in sections 5 and 6.

## 3. Evection Level Curves

Given a terrestrial spin rate and lunar semi-major axis (which define $\eta$ and $\alpha$), the Jacobi constant defines the set of allowed $(\varepsilon, \theta)$ combinations. Using the $\mathcal{O}(\varepsilon^2)$ expressions in eqns. (2.19) and (2.20), we set

$$\frac{\varepsilon}{5\alpha} = \left[\frac{\varepsilon_*}{5\alpha} + \cos 2\theta\right] \pm \left[\left(\frac{\varepsilon_*}{5\alpha} + \cos 2\theta\right)^2 + \frac{2\tilde{J}}{(5\alpha)^2}\right]^{1/2} . \qquad (3.1)$$

Figure 1 shows the resulting level curves with $x = -\sqrt{\varepsilon/5\alpha} \sin\theta$ and $y = \sqrt{\varepsilon/5\alpha} \cos\theta$ for constant $\tilde{J}/(5\alpha)^2$ values for several $\varepsilon_*/5\alpha$ values. The radial distance from the origin is equal to $\sqrt{\varepsilon/5\alpha}$ (and thus proportional to $e$), while $\theta$ is the angle from the $y$-axis in the direction of the negative $x$-axis.

The external solar torque is found from eqn. (2.1) with $T = -m\, \partial\Phi_\odot/\partial\theta$, viz.,
$$T = (15/4)m(a\Omega_\odot)^2 \varepsilon \sin 2\theta \quad;\quad T' = 10\gamma\chi\alpha a'^{1\backslash 2}\varepsilon \sin 2\theta, \qquad (3.2\text{a,b})$$
the latter being its normalized version, i.e., $T' \equiv T/(C\Omega_\oplus/t_T) = T/(C\Omega_\oplus\Omega_\odot/\chi)$. All level curves in Figure 1 have reflection symmetry across the $y$-axis, and the value of $\varepsilon$ at $\theta = \theta_0$ is equal to that at $\theta = -\theta_0$. Thus the solar torque at $\theta = \theta_0$ will be of equal magnitude but opposite sign to that at $\theta = -\theta_0$ due to the $\sin 2\theta$ term in (3.2), and the net solar torque (and thus the change in Earth-Moon AM) over a libration cycle is zero in the absence of tides.

[Figure 1]

### 3.1 Separatrix

At the stationary points $\dot{\theta} = 0$, and so from (2.21b), $\varepsilon_s = \varepsilon_{sx} = \beta$. Jacobi values at the stationary points are $\tilde{J}_s = -\varepsilon_s^2/2$ and $\tilde{J}_{sx} = -\varepsilon_{sx}^2/2$. The partial derivative of $\tilde{J}$ with respect to $\varepsilon$ is simply $\partial\tilde{J}/\partial\varepsilon = \varepsilon - \beta$ and vanishes at the stationary points, while $\partial^2\tilde{J}/\partial\varepsilon^2 = 1$ is positive, indicating a relative minimum. On the other hand, while $\partial\tilde{J}/\partial\theta = 10\alpha\varepsilon \sin 2\theta$ also vanishes, $\partial^2\tilde{J}/\partial\theta^2 = 20\alpha\varepsilon \cos 2\theta$ is positive on the $y$-axis (when $\theta = 0, \pi$) but negative on the $x$-axis (when $\theta = \pm\pi/2$). Accordingly, on the $y$-axis the stationary points are absolute minima and stable ($\varepsilon_s$; Figure 1, filled markers), while on the $x$-axis they are unstable saddle points ($\varepsilon_{sx}$; Figure 1, open markers). Note that $\tilde{J}$ is always zero at the origin and (per eqn. 2.20a) along the trajectory $\varepsilon = 2\beta$, provided that $\beta > 0$.

The level curve passing through the saddle points (Figure 1, dashed curve) is a separatrix that partitions resonant trajectories, which librate about the stable stationary points, from non-resonant trajectories that circulate around the origin. The value of $\varepsilon$ along the separatrix can be found by setting $\tilde{J} = \tilde{J}_{sx}$ in eqn. (2.20b) to give
$$\varepsilon_\pm = \varepsilon_* + 5\alpha \cos 2\theta \pm [(\varepsilon_* + 5\alpha \cos 2\theta)^2 - \varepsilon_{sx}^2]^{1/2}. \qquad (3.3)$$
where $\varepsilon_+$ ($\varepsilon_-$) denotes the radially outer (radially inner) curve. The maximum and minimum $\varepsilon_\pm$ occur at $\theta = 0, \pi/2$:
$$\varepsilon_\pm \to \varepsilon_s \pm (\varepsilon_s^2 - \varepsilon_{sx}^2)^{1/2} = \varepsilon_s \pm 2\sqrt{5\alpha\varepsilon_*}. \qquad (3.4)$$

### 3.2 Resonance domains

As $\varepsilon_*/5\alpha$ increases from initially negative (pre-capture) values to positive values, different domains emerge on the level curve diagrams. Let $\Upsilon_1$ refer to the domain area where the level curves circulate the origin in the counter-clockwise direction. When $\varepsilon_*/5\alpha < -1$, this is the only domain that exists (Figure 1a). This is *pre-capture* behavior where $\tilde{J}$ must be positive because $\beta < 0$ in this domain. In this stage both $\varepsilon_s$ and $\varepsilon_{sx}$ are negative and so $e_s$ and $e_{sx}$ are undefined.

A smaller $s'$ and/or larger $a'$ increases $\varepsilon_*$. When $-1 < \varepsilon_*/5\alpha < 1$, the stable stationary points $\varepsilon_s$ first appear at the origin (Figure 1b). With increasing $\varepsilon_*$, the stationary points $\varepsilon_s$ move outward along the $y$-axis (Figures 1c-1d). The Jacobi constant can then take on negative values down to $\tilde{J}_s = -\varepsilon_s^2/2$, which is an absolute minimum, while the level curve for $\tilde{J} = 0$, *viz.*, $\varepsilon = 2\beta$, becomes a boundary that separates trajectories that still circulate the origin (in domain $Y_1$) from a new class that librate about the stable stationary point within a new domain $Y_2$. We refer to this initial stage in resonance in which only domains $Y_1$ and $Y_2$ exist as *shallow resonance*.

For $\varepsilon_*/5\alpha > 1$, the minimum value of $\tilde{J}$ along the x-axis is no longer at the origin but occurs at new stationary points at $(\varepsilon, \theta) = (\varepsilon_{sx}, -\pi/2)$ and $(\varepsilon_{sx}, \pi/2)$. These are the saddle points where the two branches of the $\tilde{J}_{sx} = -\varepsilon_{sx}^2/2$ curve connect[2]. Trajectories for $\tilde{J} < \tilde{J}_{sx}$ still librate about the stationary points on the y-axis in domain $Y_2$. For $\tilde{J} > \tilde{J}_{sx}$, trajectories beyond the outer separatrix boundary circulate the origin counterclockwise, but within the lower separatrix boundary, there is now a new, lens-shaped domain $Y_3$, where non-resonant trajectories circulate the origin in a clockwise sense (Figures 1e-1f). We refer to this stage as *deep resonance*, whose structure above the x-axis is illustrated schematically in Figure 2.

[Figure 2]

## 4. Tidal Friction

The level curve patterns are set by the Earth's spin rate and the lunar semi major axis through $\varepsilon_*$ and $\alpha$, which evolve due to tidal friction between the Earth and Moon.

Earth-Moon tides exchange AM between the objects' spins and the lunar orbit, but do not change the total Earth-Moon AM. We represent the semi-major axis and eccentricity rates of change due to tides raised on the Earth by the Moon as $\dot{a}'_\oplus$ and $\dot{\varepsilon}_\oplus$, while $\dot{a}'_m$ and $\dot{\varepsilon}_m$ denote corresponding rates for tides raised on the Moon by the Earth. Tides alter the respective spins of the Earth and Moon at rates

$$\dot{s}' = -(\gamma/2) a'^{1/2} (1-\varepsilon)^{1/2} [\dot{a}'_\oplus/a' - \dot{\varepsilon}_\oplus/(1-\varepsilon)] \tag{4.1}$$
$$\dot{s}'_m = -(\gamma/2\kappa) a'^{1/2} (1-\varepsilon)^{1/2} [\dot{a}'_m/a' - \dot{\varepsilon}_m/(1-\varepsilon)] \tag{4.2}$$

In these expressions, the time derivatives use the afore mentioned time variable $\tau = t/t_T$, to be specified below. Conservation requires that the change in the Moon's orbital AM due to tides is $\dot{L}'_{orb,T} = -\dot{s}' - \kappa \dot{s}'_m$, or

$$\dot{L}'_{orb,T} = L'_{orb}[\dot{a}'/a' - \dot{\varepsilon}_T/(1-\varepsilon)]/2 \tag{4.3}$$

where $\dot{a}' = \dot{a}'_\oplus + \dot{a}'_m$ and $\dot{\varepsilon}_T = \dot{\varepsilon}_\oplus + \dot{\varepsilon}_m$ are the total rates of change from both Earth and lunar tides.

The equations of motion derived in section 2.4 (eqns. 2.21a,b) must be modified to include tidal changes, with

$$\dot{\varepsilon} = -20\chi\alpha(\tau)\varepsilon \sin 2\theta + \dot{\varepsilon}_T \quad ; \quad \dot{\theta} = 2\chi[\varepsilon - \varepsilon_*(\tau) - 5\alpha(\tau)\cos 2\theta]. \tag{4.4a,b}$$

As a result, a time independent first integral (Jacobi constant) no longer exists, with

$$\dot{\tilde{J}} = (\varepsilon - \varepsilon_* - 5\alpha \cos 2\theta)\dot{\varepsilon}_T - (\dot{\varepsilon}_* + 5\dot{\alpha} \cos 2\theta)\varepsilon, \tag{4.5}$$

and system trajectories on a level curve diagram are not closed. Since both $\varepsilon_*$ and $\alpha$ vary with time, so do $\varepsilon_s$ and $\varepsilon_{sx}$, although on a tidal timescale much longer then a libration period, *i.e.*, they are quasi-stationary states.

---

[2] There are now two locales where $\tilde{J} = 0$, a circulating track: $\varepsilon = 2\beta$, in domain $Y_1$ beyond the outer separatrix branch and the origin, which has switched to a local maximum.

### 4.1 Mignard Tidal Model

We employ the model of Mignard (1980), in which the rise of the equilibrium tidal distortion is delayed by a fixed time relative to the tide raising stress. We define the tidal time constant, $t_T \equiv (6k_T \mu \Omega_\oplus^2 \Delta t)^{-1}$, where $k_T$ is the Earth's 2nd degree tidal Love number and $\Delta t$ is the terrestrial time delay; for the current Earth, $\Delta t \approx 12$ min, and $t_T \sim 4 \times 10^{-4} k_T^{-1}$ years. The constant time delay results in a frequency-dependent lag-angle between the tide and the line connecting the Earth-Moon centers, $\delta = (s - n)\Delta t$, where $\delta$ varies smoothly as frequencies approach and pass through the $s = n$ case (*i.e.*, a spin synchronous with the lunar mean motion). This is a key advantage of the Mignard model compared with the Darwin-Kaula constant lag-angle tidal model (*e.g.*, Kaula, 1964), in which the lag angle has discontinuities near commensurabilities (*e.g.*, Kaula, 1964; Tian *et al.*, 2017).

*4.1.1. Earth Tides.* Considering the second harmonic in the tidal potential, the Mignard equations for the evolution of $a'$ and $\varepsilon$ vs. $\tau = t/t_T$ due to Earth tides are

$$\dot{a}'_\oplus/a' = (1 + \mu)\left[s' a'^{3/2} f_1(\varepsilon) - f_2(\varepsilon)\right]/a'^8 \tag{4.6a}$$

$$\dot{\varepsilon}_\oplus = (1 + \mu)\varepsilon\left[s' a'^{3/2} g_1(\varepsilon) - g_2(\varepsilon)\right]/a'^8 \tag{4.6b}$$

with $f_1, f_2, g_1$ and $g_2$ given by:

$$f_1(\varepsilon) = \tilde{f}_1(\varepsilon)/(1-\varepsilon)^6 \quad ; \quad f_2(\varepsilon) = \tilde{f}_2(\varepsilon)/(1-\varepsilon)^{15/2} \tag{4.7a,b}$$

$$g_1(\varepsilon) = \tilde{g}_1(\varepsilon)/(1-\varepsilon)^5 \quad ; \quad g_2(\varepsilon) = \tilde{g}_2(\varepsilon)/(1-\varepsilon)^{13/2} \tag{4.7c,d}$$

where $\tilde{f}_1$, $\tilde{f}_2$, $\tilde{g}_1$, and $\tilde{g}_2$ are polynomials in $\varepsilon$ (Table 2) found by orbit averaging the tidal forces. Combining these with eqn. (4.1), the de-spin rate of the Earth is

$$\dot{s}' = -\frac{\gamma(1+\mu)}{2a'^{15/2}}\left(s'a'^{3/2}\frac{\tilde{f}_1 - \varepsilon\tilde{g}_1}{(1-\varepsilon)^{11/2}} - \frac{\tilde{f}_2 - \varepsilon\tilde{g}_2}{(1-\varepsilon)^7}\right) \tag{4.8}$$

[Table 2]

*4.1.2 Lunar Tides.* The corresponding evolution expressions due to satellite tides are

$$\dot{a}'_m/a' = (1 + \mu)A\left[s'_m a'^{3/2} f_1(\varepsilon) - f_2(\varepsilon)\right]/a'^8 \tag{4.9a}$$

$$\dot{\varepsilon}_m = (1 + \mu)A\varepsilon\left[s'_m a'^{3/2} g_1(\varepsilon) - g_2(\varepsilon)\right]/a'^8 \tag{4.9b}$$

where

$$A \equiv \left(\frac{k_m}{k_T}\right)\left(\frac{\Delta t_m}{\Delta t}\right)\left(\frac{M}{m}\right)^2\left(\frac{R_m}{R}\right)^5 \approx 10\left(\frac{k_m}{k_T}\right)\left(\frac{\Delta t_m}{\Delta t}\right) \tag{4.10}$$

is a ratio of physical parameters of the two bodies that scales the relative strength of tides on the Moon to tides on the Earth, with $R_m, k_m$, and $\Delta t_m$ referring to the Moon's radius, tidal Love number and tidal time delay (Mignard, 1980). For the current Earth and Moon, $A \approx$ unity. However, when the early Moon encountered evection, the post-giant impact Earth would have still been fully molten, with a tidal response akin to that of a fluid body with a small $\Delta t$, while the Moon would have likely cooled sufficiently to yield a dissipative state with a much larger $\Delta t_m$, implying $A \gg 1$ when the Moon encountered the resonance (Zahnle *et al.*, 2015).

Combining eqns. (4.9a-b) and (4.10) with eqn. (4.2), the change in the lunar spin rate is

$$\dot{s}'_m = -\frac{\gamma}{2\kappa}\frac{A(1+\mu)}{a'^{15/2}}\left[s'_m a^{3/2}\frac{\tilde{f}_1 - \varepsilon\tilde{g}_1}{(1-\varepsilon)^{11/2}} - \frac{\tilde{f}_2 - \varepsilon\tilde{g}_2}{(1-\varepsilon)^7}\right] \tag{4.11}$$

Because the Moon-Earth mass ratio $\mu$ is small, we set $(1 + \mu) \approx 1$ in all subsequent tidal rate expressions.

*4.1.3 Lunar rotation.* For a synchronously rotating satellite, $s'_m a'^{3/2} = 1$, and the above equations would simplify to $\dot{a}'_m/a' = A(f_1 - f_2)/a'^8$, $\dot{\varepsilon}_m = A\varepsilon(g_1 - g_2)/a'^8$, and $\dot{s}'_m = 0$. However, there is a contradiction here. For an eccentric orbit with $\varepsilon > 0$, eqn. (4.2) shows that the satellite's spin will not remain at a constant value of $s'_m = a'^{-3/2}$ if subject to tidal rates given in eqn. (4.9a,b). Instead there will be a non-zero torque on the satellite spin that will move it away from synchronicity until

$$s'_m a'^{3/2} = \frac{(1-\varepsilon)f_2 - \varepsilon g_2}{(1-\varepsilon)f_1 - \varepsilon g_1}, \tag{4.12}$$

which is only unity for a circular orbit ($\varepsilon = 0$), implying non-synchronous rotation for an eccentrically orbiting satellite. Synchronous rotation can be maintained in an eccentric orbit if an additional torque is exerted on a permanent triaxial figure of the Moon (*e.g.*, Goldreich, 1966; Goldreich and Peale, 1966a,b; Aharonson *et al.*, 2012). The original Mignard equations that assumed synchronous rotation did not include this permanent figure torque. Appendix D develops expressions to include this torque's effects on $a$ and $\varepsilon$ for cases in which synchronous lunar rotation is assumed.

4.2 Resonance Encounter

During the initial pre-capture expansion of the lunar orbit, $\varepsilon_*$ is negative but increasing. Shallow resonance is first established when $\varepsilon_* = -5\alpha$ and $s' = a'^{7/4}(1 + 4\alpha)/\Lambda$. Assuming that prior to that the Moon's orbit was circular, its spin synchronous, and the system angular momentum, $L'_o$, conserved, yields the constraint

$$L'_o = (1 + 4\alpha_o a'^{3/2}_{res})a'^{7/4}_{res}/\Lambda + \kappa/a'^{3/2}_{res} + \gamma a'^{1/2}_{res} \tag{4.13}$$

for the resonance encounter distance as displayed in Figure 3. For $L'_o$ equal to the current system value, $L'_{EM} = 0.346$, $a'_{res} = 4.61$ and $s'_{res} = 0.267$. For a high-AM state with $L'_o \approx 2L'_{EM}$, one obtains $a'_{res} = 7.30$ and $s'_{res} = 0.596$. For a low-AM state with $L'_o < 0.190$ ($0.549 L_{EM}$), evection would lie interior to the Roche limit and would not be encountered as the Moon's orbit tidally expanded.

[Figure 3]

Capture into resonance requires that tidally driven changes in the stationary point occur slowly compared to the resonant libration timescale. To understand the condition required to maintain the resonance, we differentiate eqn. (2.21b) (in the limit of no tides) and then use both (2.21a,b) to eliminate $\dot{\varepsilon}$ and $\dot{\theta}$ to yield,

$$\ddot{\theta} = -40\chi^2 \alpha(\varepsilon_* + 5\alpha \cos 2\theta) \sin 2\theta \approx -80\chi^2 \alpha \varepsilon_s \theta \tag{4.14}$$

where the final version assumes small $\theta$. This is the equation for a harmonic oscillator of frequency $\omega = 4\chi\sqrt{5\alpha\varepsilon_s}$ that is librating about a stable equilibrium point. The libration frequency increases with $\varepsilon_s$. When in the shallow resonant regime, if the time it takes to execute a half cycle around the stationary point, $\sim\pi/\omega$, is comparable to or shorter than the time for that point to reach a given $\varepsilon_s$ value via tides, $\sim\varepsilon_s/|\dot{\varepsilon}_s|$, capture into region $Y_2$ can occur. This condition requires that $\varepsilon_s \geq [\pi\dot{\varepsilon}_s/(4\chi\sqrt{5\alpha})]^{2/3}$. For slow tides (small $\dot{\varepsilon}_s$), this can be satisfied for small $\varepsilon_s$, but the needed $\varepsilon_s$ value increases for faster tidal evolution. On the other hand, once $\varepsilon_s \geq 10\alpha$ (*i.e.*, once $\varepsilon_*/5\alpha \geq 1$) the saddle points, $\varepsilon_{sx}$, appear, and an increasing portion of phase space becomes occupied by the inner non-resonant region $Y_3$ (*e.g.*, Figure 2), causing the resonant region $Y_2$ to radially narrow.

This makes the adiabatic condition for resonance stability more stringent as $\varepsilon_*/5\alpha$ increases further, because smaller tidally-driven changes in $\varepsilon_s$ during a libration cycle can cause the trajectory to pass directly from $Y_1$ to $Y_3$, avoiding resonance capture.

## 5. Evolution: Damped Libration

We first construct a baseline evolutionary track by restricting the Moon's resonance behavior to one of zero libration amplitude, for which the eccentricity equals that of the stable stationary state (*i.e.,* we set $\varepsilon = \varepsilon_s$, $\theta = 0$, and ignore eqns. (4.4a,b) associated with libration about the stationary state). While obviously idealized, the damped libration solution reveals how angular momentum drain occurs and when in the evolution it would be most significant if the resonance is maintained. In Section 6, we expand on this baseline evolution to estimate when libration amplitude growth and resonance escape is expected.

It is uncertain whether the Moon would have had a permanent triaxial moment when it encountered evection. Our nominal damped libration cases consider a non-synchronously rotating moon without a permanent figure. An example case assuming a triaxial moon in synchronous rotation is presented in Appendix E (Figure A3). For non-synchronous cases, we assume that the lunar spin state rapidly reaches the steady state value from (4.12). That $s'_m$ would, in the absence of permanent figure torques, rapidly reach this value can be seen from eqns. (4.8) and (4.11), where for an initial $s'_m \sim s'$, the rate of change of the lunar spin is larger than the rate of change of the Earth's spin by a factor of $A/\kappa$, which is $\geq 10^3$ for $A \geq 1$.

### 5.1. Angular Momentum Loss.

To estimate the rate at which the evection resonance could drain angular momentum from the Earth-Moon, eqn. (2.7) is differentiated with respect to time,

$$\dot{L}' = \dot{s}' + \kappa \dot{s}'_m + \dot{L}_{orb} = \dot{s}' + \kappa \dot{s}'_m + (\gamma/2) a'^{1/2} (1-\varepsilon_s)^{1/2} \left(\frac{\dot{a}'}{a'} - \frac{\dot{\varepsilon}_s}{1-\varepsilon_s}\right). \quad (5.1)$$

Note that the R.H.S. applies only once $\varepsilon_s \geq 0$ (post-resonance capture), because for $\varepsilon_s < 0$ (pre-capture), the stationary eccentricity is undefined. Spin rates and the Moon's semi-major axis are affected only by tides, and since tides alone would conserve system angular momentum, it follows that

$$\dot{s}' + \kappa \dot{s}'_m = -(\gamma/2) a'^{1/2} (1-\varepsilon_s)^{1/2} \left(\frac{\dot{a}'}{a'} - \frac{\dot{\varepsilon}_T}{1-\varepsilon_s}\right) = -\dot{L}_{orb,T}. \quad (5.2)$$

Substituting into (5.1), we confirm that,

$$\dot{L}' = \dot{L}_{orb} - \dot{L}_{orb,T} = (\gamma/2) a'^{1/2} (\dot{\varepsilon}_T - \dot{\varepsilon}_s)/(1-\varepsilon_s)^{1/2}, \quad (5.3)$$

which again applies only once $\varepsilon_s \geq 0$. Thus the change in angular momentum reflects the difference between the rate at which tides change the lunar orbit eccentricity vs. the rate of change of eccentricity imposed by evection. The rate due to tides, $\dot{\varepsilon}_T = \dot{\varepsilon}_\oplus + \dot{\varepsilon}_m$, is given by eqns. (4.6b) and (4.9b), whereas $\dot{\varepsilon}_s$ can be found by differentiating eqn. (2.16),

$$\dot{\varepsilon}_s \approx \eta \left(\frac{7}{4} \frac{\dot{a}'}{a'} - \frac{\dot{s}'}{s'}\right) + 6\alpha \frac{\dot{a}'}{a'} \quad (5.4)$$

The latter rate is determined primarily by the tidal changes of $a'$ and $s'$, instead of $\dot{\varepsilon}_T$.[3] The above utilizes the expansion to $\mathcal{O}(\varepsilon^2)$ from section 2.4; however including higher order terms does not substantially alter the overall behavior so long as $\alpha$ is small.

---

[3]Although $\dot{s}'$ does depend on $\dot{\varepsilon}_\oplus$ through eqn. (4.1).

### 5.2. Tidal Evolution in Resonance with no Libration

Figure 4 illustrates a zero-libration evolution for an initial angular momentum $L_o = 2L_{EM}$ and $A = 10$. Additional evolutions for varied $L_o$ and $A$ values are presented in Appendix E (Figures A1 and A2). Pre-resonance capture, the semi-major axis (Figure 4a) grows at a rate that decreases with distance, while the eccentricity (Figure 4b) remains zero and the total angular momentum constant (Figure 4d). Once the Moon's orbit is captured in evection ($\varepsilon_s \geq 0$), its eccentricity rises (gray regions in Figured 4a-b,d), decreasing $L'_{orb}$ somewhat even though the outward migration temporarily speeds back up. The eccentricity eventually reaches a critical value, $\varepsilon_c$, at which outward orbit migration stalls and the orbit begins to contract due to the effect of satellites tides. Soon after, the eccentricity begins to decline as well[4], and the Moon enters a prolonged contraction phase. If the resonance is maintained with zero libration throughout the evolution, the system would ultimately reach a co-synchronous end-state, $s = s_m = n$, with zero eccentricity. However, libration amplitude growth and escape from resonance is predicted long before that state is achieved (see Section 6).

[Figure 4]

The rates $\dot{a}'/a$, $\dot{\varepsilon}_T$ and $\dot{\varepsilon}_s$ during the evolution in Figure 4 are shown in Figure 5a, while 5b displays $\dot{s}'$, $\dot{L}'_{orb}$ and $\dot{L}'$. The slowdown of the Earth's spin continues throughout the evolution. The maximum decay rate of $L'_{orb}$ occurs near the start of lunar contraction, but quickly diminishes to a small value. Throughout the rest of the evolution, the orbital AM remains relatively constant in spite of continued changes in $a'$ and $\varepsilon_s$. As a result, $\dot{L}'$ and $\dot{s}'$ become nearly equal (eqn. 5.1), i.e., the AM drained from the system by evection is nearly completely reflected in the concomitant slowdown in the Earth's spin. We now examine each evolutionary stage in greater detail.

[Figure 5]

*5.2.1 Outward migration.* For a Moon in an initially circular orbit outside the Earth's co-rotation radius $\dot{s}'$ is negative, while the lunar orbital angular momentum, $L'_{orb}$, increases to compensate so that $dL'/d\tau = 0$. After resonance capture the Moon's orbit continues to expand due to tides ($\dot{a}' > 0$), while evection increases the Moon's eccentricity ($\dot{\varepsilon}_s > 0$; gray area Figure 4) per eqn. (5.4). Concentrating on just the rate of change of the *orbital* angular momentum, $\dot{L}'_{orb}$, given by the last term of (5.1) once $\varepsilon_s > 0$, we can write

$$\dot{L}'_{orb} = (L'_{orb}/2)\left[\dot{a}'/a' - \dot{\varepsilon}_s/(1 - \varepsilon_s)\right] \approx (L'_{orb}/2)\left[-3\dot{a}'/4a' + \dot{s}'/s'\right] \quad (5.5)$$

Both terms in the final bracket are negative during this phase, and the angular momentum of the Moon's orbit decreases with time (Figure 4d) even though its semi-major is increasing.

As evection increases the Moon's eccentricity, it eventually reaches a critical value, $\varepsilon_c$, at which there is a balance between the rates at which Earth and lunar tides alter the Moon's semi-major axis ($\dot{a}'_\oplus \approx -\dot{a}'_m$), and the Moon's orbital expansion stalls at $a' = a'_c$. If the Moon had a very small eccentricity when first captured into resonance at $a'_{res}$, the change in its orbital angular momentum during its migration from there to $a'_c$ would be $\Delta L'_{orb,evec} = \gamma[a'^{1/2}_c(1 - \varepsilon_c)^{1/2} - a'^{1/2}_{res}]$. For very small initial eccentricity, $\eta = \Lambda s'/a'^{7/4} \approx 1$ and the Earth's spin upon capture is $a'^{7/5}_{res}/\Lambda$, while at $a' = a'_c$, $\eta = \Lambda s'_c/a'^{7/4}_c \approx 1 - \varepsilon_c$ (neglecting small terms proportional to $\alpha_0$) so that the corresponding change in the Earth's spin angular momentum is $\Delta s'_{evec} \approx [a'^{7/4}_c(1 - \varepsilon_c) - a'^{7/4}_{res}]/\Lambda$. By comparison, the changes in the absence of evection would be simply $\Delta L'_{orb,T} = -\Delta s'_T = \gamma(a'^{1/2}_c - a'^{1/2}_{res})$. For the $A = 10$, $L_o = 2L_{EM}$ case shown in Figure 4, $a'_{res} \approx 7.30$, $a'_c \approx 9.8$, $\varepsilon_c \approx 0.43$ and we find, $\Delta L'_{orb,evec} \approx -0.012$, $\Delta s'_{evec} \approx -0.027$, for a total loss of $\Delta L'_{evec} = -0.039$. Compared to the initial angular momentum of the Earth-Moon system in this case ($L'_0 = 0.69$), this is only a modest, $\sim 6\%$ reduction. In general, if evection is active only during the Moon's outbound phase, as was found by Touma and Wisdom (1998), the resulting angular momentum change is small, consistent with prior assumptions of canonical giant impact models (*e.g.*, Canup, 2008).

---
[4]The times at which $\dot{a}$ and $\dot{e}$ vanish are slightly different.

*5.2.2 Inward migration.* Subsequent to stalling, the lunar orbit begins to contract. Provided the resonance condition is maintained and evection continues to control the Moon's eccentricity (as assumed in the zero libration evolution here), $\varepsilon$ soon begins to decrease as well. Earth tides further drain $s$ as long as, $(1 - \varepsilon_s)^{-1}\dot\varepsilon_s < \dot a'_\oplus/a'$ (eqn. 4.1). It is during this secondary contraction phase that substantial angular momentum may be lost, as was seen in the simulations of CS12.

The contraction of the lunar orbit occurs relatively slowly because the magnitude of $\dot a'_m < 0$ due to lunar tides is only somewhat larger than the opposing action $\dot a'_\oplus$ due to Earth tides. The result is a prolonged period during which angular momentum that is removed from the Earth's spin by Earth tides can be transferred by the resonance to the Earth's orbit, with $\dot s' \sim T'$ (as seen in Figure 5b where $\dot s' \sim \dot L'$). The changes in the components of the Earth-Moon angular momentum during this phase would be $\Delta L'_{orb}|_{evec} = \gamma[a'^{1/2}_{esc}(1 - \varepsilon_{esc})^{1/2} - a'^{1/2}_c(1 - \varepsilon_c)^{1/2}]$ and $\Delta s'_{evec} = [a'^{7/4}_{esc}(1 - \varepsilon_{esc}) - a'^{7/4}_c(1 - \varepsilon_c)]/\Lambda$, where $a'_{esc}$, $\varepsilon_{esc}$ are the semi-major axis and eccentricity (squared) at the time of resonance escape. For the particular evolution shown in Figure 4 (with $A = 10, L_0 = 2L_{EM}$), the system angular momentum decreases to that of the current Earth-Moon system ($L'_{EM} = 0.35$) at Time/$t_T = 1.4 \times 10^6$. For resonance escape to occur at this point implies $a'_{esc} = 4.7$ and $\varepsilon_{esc} = 0.06$.

From Figure 4d (and also Figures A1-A3 in Appendix E), it can be seen that the longer the Moon remains in resonance, the greater the reduction in $L$, so that the final angular momentum achieved via formal evection will be set by the timing of resonance escape. Escape can occur if the adiabatic condition is violated (so that the timescale for tidally driven changes in $\varepsilon$ becomes short compared to the resonant libration timescale), or if the libration amplitude grows and exceeds the maximum amplitude of $\pi/2$ consistent with resonant libration. If escape never occurred, evection would drain the system's angular momentum until the dual synchronous state is achieved. The limiting final angular momentum in this case is found by setting $s'$, $s'_m = s'_{sync} = a'^{-3/2}$ and $\varepsilon_s = 0$, *viz.*,

$$L'_{sync} = (1 + \kappa)/a'^{3/2}_{sync} + \gamma a'^{1/2}_{sync} \qquad (5.6)$$

where $a'_{sync}$ is the final semi-major axis. This will be right at the inner boundary of the resonance where $\varepsilon_s$ can go to zero, and is obtained by setting $\Lambda s'_{sync}/a'^{7/4}_{sync} = \eta_{sync} = 1 + 4\alpha_{sync}$, and then solving for $a'_{sync} \cong \Lambda^{4/13} = 3.416$. Eqn. (5.6) then gives $L'_{sync} = 0.226$, which is substantially less than that of the current Earth-Moon ($L'_{EM} = 0.35$). In addition to being inconsistent with the Earth-Moon AM, a dual synchronous state would also be unstable, because further slowing of the Earth's spin by direct solar tides would eventually cause synchronous orbit to drift beyond the Moon, which would then tidally evolve inward. Clearly this full evolution in evection never occurred for the Earth-Moon pair, and indeed in section 6 we predict much earlier resonance escape.

### 5.3 Tidal Stationary States

The above baseline evolution adopts the stationary state eccentricity in the absence of tides. If tides are included as in eqns. (4.4a,b), eqn. (4.14) describing libration about the stationary state is replaced by

$$\ddot{\theta} = -40\chi^2\alpha(\varepsilon_* + 5\alpha\cos 2\theta)\sin 2\theta + 2\chi(\dot{\varepsilon}_T - \dot{\varepsilon}_* - 5\dot{\alpha}\cos 2\theta) \qquad (5.7)$$

and the angle, $\theta_s$, for which $\ddot{\theta}$ vanishes satisfies,

$$20\chi\alpha(\varepsilon_* + 5\alpha\cos 2\theta_s)\sin 2\theta_s = \dot{\varepsilon}_T - \dot{\varepsilon}_* - 5\dot{\alpha}\cos 2\theta_s \ . \qquad (5.8)$$

The angle $\theta_s$ represents an offset from the $y$-axis of the stationary state around which stable libration occurs that is due to the effects of tides. If the tidal rates are small enough that the offset angle is small and $\cos 2\theta_s \sim 1$, $\sin 2\theta_s \approx (\dot{\varepsilon}_T - \dot{\varepsilon}_* - 5\dot{\alpha})/20\chi\alpha(\varepsilon_* + 5\alpha)$; if instead tidal rates are fast and the offset is large so that $\cos 2\theta_s$ is small, $\sin 2\theta_s \approx (\dot{\varepsilon}_T - \dot{\varepsilon}_*)/20\chi\alpha\varepsilon_*$. However, since the maximum value of $|\sin 2\theta_s|$ is unity when $\theta_s = \pm 45°$, there can be no stationary angle (and thus no stable libration) if $|\dot{\varepsilon}_T - \dot{\varepsilon}_*| > 20\chi\alpha\varepsilon_*$.

Using the low $\cos 2\theta_s$ approximation and neglecting terms proportional to $\alpha$ and $\dot{\alpha}$, eqn. (4.4a) becomes $\dot{\varepsilon} \approx -(\dot{\varepsilon}_T - \dot{\varepsilon}_*)\varepsilon/\varepsilon_* + \dot{\varepsilon}_T$, and differentiating yields

$$\ddot{\varepsilon} \approx [-\dot{\varepsilon} + (\dot{\varepsilon}_*/\varepsilon_*)\varepsilon](\dot{\varepsilon}_T - \dot{\varepsilon}_*)/\varepsilon_* + (1 - \varepsilon/\varepsilon_*)\ddot{\varepsilon}_T + (\varepsilon/\varepsilon_*)\ddot{\varepsilon}_* \ . \qquad (5.9a)$$

Ignoring $\ddot{\varepsilon}_T$, $\ddot{\varepsilon}_*$, using the $\dot{\varepsilon}$ expression above eqn. (5.9a) and requiring $\ddot{\varepsilon} \to 0$ to suppress oscillations, results in

$$(\varepsilon/\varepsilon_* - 1)(\dot{\varepsilon}_T - \dot{\varepsilon}_*)\dot{\varepsilon}_T/\varepsilon_* \approx 0 \qquad (5.9b)$$

Assuming non-zero tidal rates ($\dot{\varepsilon}_T \neq 0$), satisfying this condition implies $\varepsilon_s \approx \varepsilon_*$, vs. $\varepsilon_s \approx \varepsilon_* + 5\alpha$ found in the $\mathcal{O}(\varepsilon^2)$ expansion in section 2.4. Thus the stationary eccentricity is relatively unaffected by an increasing stationary offset angle imposed by tides. Note that substituting these state parameters into eqns. (4.4a,b) will not give zero values for $\dot{\varepsilon}_s$ and $\dot{\theta}_s$ because they are now slowly changing quasi-steady states.

## 6. Evolution: Finite Libration and Resonance Escape

Until now, the orbit evolution has been artificially constrained to zero libration. On the other hand, Touma and Wisdom (1998) found that the Moon escapes evection soon after it reaches the distance where $\dot{a} = 0$. At this point, they found that the resonance libration amplitude begins to rapidly increase until the system leaves resonance. Escapes were also reported by CS12, although much later in the evolution during the orbital contraction phase. In this section, we explore how tides affect the libration behavior, libration amplitude growth, and the expected timing of resonant escape.

At the turn-around point of a level curve, $\partial\tilde{J}/\partial\varepsilon|_\Theta = 0$. From eqn. (2.20a), this implies that

$$\varepsilon_\Theta = \varepsilon_* + 5\alpha\cos 2\Theta \qquad (6.1)$$

where $\varepsilon_\Theta$ denotes the eccentricity at turn-around and we have assumed $\dot{\alpha}$ is small during a libration cycle. Substituting into $\tilde{J}$ then leads to

$$\varepsilon_\Theta^2 = -2\tilde{J} \quad ; \quad \cos 2\Theta = (\sqrt{-2\tilde{J}} - \varepsilon_*)/5\alpha \qquad (6.2a,b)$$

### 6.1 Libration Amplitude Variation

To examine the behavior of the libration amplitude on an oscillation timescale, we wish to integrate eqn. (5.7) including eccentricity variations over a libration cycle. We consider a case where the libration amplitude is small and retain only terms linear in $\theta$ to find,

$$\ddot{\theta} + \omega^2 \theta \approx 2\chi(\dot{\varepsilon}_T - \dot{\varepsilon}_s) \equiv \mathcal{F} \tag{6.3}$$

where again $\omega^2 \equiv 80\chi^2 \alpha \varepsilon_s$. As in eqn. (4.14), this resembles a harmonic oscillator of frequency $\omega$, but now with an additional forcing term, $\mathcal{F}$, due to tides. The solution to eqn. (6.3) has two parts: a homogeneous solution, $\theta_h = \Theta \sin \omega\tau$, equal to that of the unforced equation, and a particular solution,

$$\theta_p = -\frac{\cos \omega\tau}{\omega} \int \mathcal{F}(\tau) \sin \omega\tau \, d\tau + \frac{\sin \omega\tau}{\omega} \int \mathcal{F}(\tau) \cos \omega\tau \, d\tau \tag{6.4}$$

The $\mathcal{F}(\tau)$ term has an oscillating part through its $\varepsilon$ dependence over a libration cycle and can be expanded to lowest order around its value $\mathcal{F}_s \equiv \mathcal{F}(a', s', \varepsilon_s)$ at $\varepsilon_s$, i.e.,

$$\dot{\varepsilon}_T \approx \dot{\varepsilon}_T(\varepsilon_s) + \frac{\partial \dot{\varepsilon}_T}{\partial \varepsilon}(\varepsilon - \varepsilon_s) + \cdots \quad ; \quad \dot{\varepsilon}_s \approx \dot{\varepsilon}_s(\varepsilon_s) + \frac{\partial \dot{\varepsilon}_s}{\partial \varepsilon}(\varepsilon - \varepsilon_s) + \cdots \tag{6.5a}$$

implying

$$\mathcal{F}(a', s', \varepsilon) \approx \mathcal{F}_s + \frac{\partial \mathcal{F}}{\partial \varepsilon}\bigg|_{\varepsilon_s}(\varepsilon - \varepsilon_s) + \cdots \tag{6.5b}$$

The lead term results in a particular solution, $\theta_s = \mathcal{F}_s/\omega^2$ that reduces to the tidal stationary angle of section 5.3. However, the second term produces a time-varying particular solution describing libration.

The linearized version of eqn. (2.21a) for small $\theta$ and $\varepsilon \approx \varepsilon_s$ reads $\dot{\varepsilon} \approx -40\chi\alpha\varepsilon_s\theta$, and utilizing the homogeneous solution for $\theta$ integrates to

$$\varepsilon - \varepsilon_s \approx 40\alpha\varepsilon_s(\chi/\omega)\Theta_o \cos \omega\tau \tag{6.6}$$

where $\Theta_o$ represents the libration amplitude at the start of a given cycle when $\omega\tau = -\pi/2$. Substituting eqns. (6.5b) and (6.6) into eqn. (6.4) and integrating we get the time-varying particular solution,

$$\theta_p = 20\alpha\varepsilon_s \frac{\chi}{\omega^2} \frac{\partial \mathcal{F}}{\partial \varepsilon} \Theta_o \tau \sin \omega\tau \, . \tag{6.7}$$

Combining and arranging terms gives the variation of $\theta$ with respect to the tidal stationary offset angle $\theta_s$,

$$\theta - \theta_s = \theta_h + \theta_p \approx \Theta_0 \left[1 + 20\alpha\varepsilon_s \frac{\chi}{\omega^2} \frac{\partial \mathcal{F}}{\partial \varepsilon} \tau\right] \sin \omega\tau, \tag{6.8}$$

and it is seen that libration relative to the offset angle will change with time due to the $\partial \mathcal{F}/\partial \varepsilon$ term, i.e., due to the variation in $(\dot{\varepsilon}_T - \dot{\varepsilon}_s)$ during a libration cycle due to small changes in $\varepsilon$. There is a resulting change, $\Delta\Theta = 40\pi\alpha\varepsilon_s(\chi/\omega^3)(\partial \mathcal{F}/\partial \varepsilon)\Theta_0$, in the oscillation amplitude after a complete cycle, $\Delta\tau = 2\pi/\omega$. This updated value then applies to the next cycle, etc., implying,

$$\frac{1}{\Theta}\frac{d\Theta}{d\tau} = 20\alpha\varepsilon_s \frac{\chi}{\omega^2}\frac{\partial \mathcal{F}}{\partial \varepsilon} = 40\alpha\varepsilon_s \left(\frac{\chi}{\omega}\right)^2 \left(\frac{\partial \dot{\varepsilon}_T}{\partial \varepsilon} - \frac{\partial \dot{\varepsilon}_s}{\partial \varepsilon}\right) = \frac{1}{2}\left(\frac{\partial \dot{\varepsilon}_T}{\partial \varepsilon} - \frac{\partial \dot{\varepsilon}_s}{\partial \varepsilon}\right) \tag{6.9}$$

where the last step uses the $\omega$ definition.

Thus whether the libration amplitude grows or damps depends on the sign of $(\partial\dot{\varepsilon}_T/\partial\varepsilon - \partial\dot{\varepsilon}_s/\partial\varepsilon)$. For example, if both $\partial\dot{\varepsilon}_T/\partial\varepsilon$ and $\partial\dot{\varepsilon}_s/\partial\varepsilon$ are positive (as occurs during the initial phase of expansion in resonance, see Figure 6a), then the libration amplitude will damp if the rate of change in the Moon's eccentricity due to tides increases more slowly with $e$ (i.e., $\varepsilon$) than does the rate of change of the stationary eccentricity. Conversely, once $\partial\dot{\varepsilon}_T/\partial\varepsilon > \partial\dot{\varepsilon}_s/\partial\varepsilon$ (which occurs near the stagnation point, $\dot{a} = 0$), the libration amplitude increases with time (e.g., Figure 6a).

The partial derivatives depend on the specific tidal model employed. For the model of Mignard (1980) used here

$$a'^8 \frac{\partial \dot{\varepsilon}_T}{\partial \varepsilon} = (s' + As'_m)a'^{3/2}\left(1 + \frac{\varepsilon}{\tilde{g}_1}\frac{\partial \tilde{g}_1}{\partial \varepsilon} + \frac{5\varepsilon}{1-\varepsilon}\right)g_1 - (1+A)\left(1 + \frac{\varepsilon}{\tilde{g}_2}\frac{\partial \tilde{g}_2}{\partial \varepsilon} + \frac{13\varepsilon/2}{1-\varepsilon}\right)g_2 \quad (6.10)$$

$$\frac{\partial \dot{\varepsilon}_s}{\partial \varepsilon} \approx \left(\frac{7\eta}{4} + 6\alpha\right)\frac{\partial}{\partial \varepsilon}\left(\frac{\dot{a}'}{a'}\right) - \frac{\eta}{s'}\frac{\partial \dot{s}'}{\partial \varepsilon}, \quad (6.11)$$

while taking the derivatives of $\dot{a}'/a'$ and $\dot{s}'$ gives

$$a'^8 \frac{\partial}{\partial \varepsilon}\left(\frac{\dot{a}'}{a'}\right) = (s' + As'_m)a'^{3/2}\left(\frac{1}{\tilde{f}_1}\frac{\partial \tilde{f}_1}{\partial \varepsilon} + \frac{6}{1-\varepsilon}\right)f_1 - (1+A)\left(\frac{1}{\tilde{f}_2}\frac{\partial \tilde{f}_2}{\partial \varepsilon} + \frac{15/2}{1-\varepsilon}\right)f_2 \quad (6.12)$$

$$\frac{\partial \dot{s}'}{\partial \varepsilon} = -\frac{1}{2}\gamma a'^{1/2}\frac{\partial}{\partial \varepsilon}\left((1-\varepsilon)^{1/2}\frac{\dot{a}'_\oplus}{a'} - \frac{1}{(1-\varepsilon)^{1/2}}\dot{\varepsilon}_\oplus\right)$$

$$= \frac{1}{2}\gamma a'^{1/2}\left(\frac{1}{2(1-\varepsilon)^{1/2}}\left(\frac{\dot{a}'_\oplus}{a'} + \frac{\dot{\varepsilon}_\oplus}{1-\varepsilon}\right) - (1-\varepsilon)^{1/2}\left(\frac{\partial}{\partial \varepsilon}\frac{\dot{a}'_\oplus}{a'} - \frac{1}{1-\varepsilon}\frac{\partial \dot{\varepsilon}_\oplus}{\partial \varepsilon}\right)\right) \quad (6.13)$$

where the $f$ and $g$ polynomials and their derivatives (Table 2) are to be evaluated for $\varepsilon = \varepsilon_s$. Setting $A = 0$ in equations (6.10) and (6.12) provides the Earth-only tidal expressions needed for $\partial \dot{s}'/\partial \varepsilon$. Analogous expressions for the case of synchronous lunar rotation maintained by a permanent figure torque are provided in Appendix D.

Figure 6 displays the partial derivative behaviors and $\Theta^{-1}d\Theta/d\tau$ for the baseline evolution shown in Figure 4 (with $A = 10$, $L_o = 2L_{EM}$). Figure 7 shows $\Theta^{-1}d\Theta/d\tau$ for varied $A$ values for $L_o = 2L_{EM}$, and for varied $L_o$ with $A = 10$, all for a non-synchronously rotating Moon. Figure 8 contrasts $\Theta^{-1}d\Theta/d\tau$ for synchronous vs. non-synchronous rotation cases, both with $A = 10$ and $L_o = 2L_{EM}$. Across all parameter choices, libration amplitude growth (*i.e.*, $\Theta^{-1}d\Theta/d\tau > 0$) is predicted for Mignard tides during the lunar orbital contraction phase.

[Figure 6]

[Figure 7]

[Figure 8]

Before estimating when libration amplitude growth would lead to resonance escape with Mignard tides (Section 6.2 below), we briefly consider application of eqn. (6.9) to the constant lag angle/constant-$Q$ tidal model utilized in Wisdom and Tian (2015). Figure 9 shows the predicted behavior of $\Theta^{-1}d\Theta/dt$ for a baseline evolution (*i.e.*, with $\varepsilon = \varepsilon_s$ and $\theta = 0$) that adopts the tidal expressions for a synchronously rotating Moon as given in Wisdom and Tian's equations (21) through (40), with $A$ now defined in their eqn. (12). It can be seen that for the $A = 1.7$ and 2.0 cases (light blue curves in Fig. 9), eqn. (6.9) predicts an extended period of libration amplitude damping that persists even as the Moon's orbit contracts, implying resonance stability. This is consistent with protracted resonance occupancy, decreasing libration amplitude, and large AM modification seen for these $A$ values in both the simplified models and full integrations of Wisdom and Tian (*e.g.*, their Figs. 2, 3, 5, and 9). However, outside this narrow range of $A$, eqn. (6.9) predicts libration amplitude excitation even prior to lunar orbit contraction (darker blue curves in Fig. 9), suggesting limited resonance stability. For this regime, Wisdom and Tian indeed found minimal or no formal resonance occupancy with constant-$Q$ tides.

Predictions from the idealized solutions developed here thus appear qualitatively consistent with results of more complete integrations with regards to formal resonance occupancy (although our methods do not allow us to assess the non-librating "limit cycle" behavior seen in Wisdom and Tian, a point we return to in Section 7). That constant-$Q$ model evolutions find prolonged damped libration in resonance for a narrow range of $A$ values (Figure 9; Wisdom and Tian, 2015), while evolutions with Mignard tides do not (Figures 7-8 and Section 6.2), thus appears to be due to differences in the tidal models themselves rather than to other differences between this work and that of Wisdom and Tian (*e.g.*, different evolution methods, inclusion of finite terrestrial obliquity and/or lunar inclination in their integrations, etc.). As the Moon's orbit contracts, $\partial \dot{\varepsilon}_T / \partial \varepsilon$ and $\partial \dot{\varepsilon}_s / \partial \varepsilon$ are negative for both the constant-$Q$ and Mignard models. However for constant-$Q$ tides with $A$ = 1.7 and 2.0, $|\partial \dot{\varepsilon}_T / \partial \varepsilon| > |\partial \dot{\varepsilon}_s / \partial \varepsilon|$ for an extended period during orbit contraction, implying damping, while for Mignard tides $|\partial \dot{\varepsilon}_T / \partial \varepsilon| < |\partial \dot{\varepsilon}_s / \partial \varepsilon|$, implying excitation (*e.g.*, Figure 6a). Beyond this narrow range of $A$, both the constant-$Q$ and Mignard models have $|\partial \dot{\varepsilon}_T / \partial \varepsilon| < |\partial \dot{\varepsilon}_s / \partial \varepsilon|$ during contraction, implying libration amplitude excitation. Differences in evolution rates (*i.e.*, $de/dt$, $da/dt$) between the constant-$Q$ and Mignard tidal models are most pronounced for high-eccentricity orbits, and so it is not surprising that the divergent outcomes occur for low $A$ cases in which the peak eccentricities are highest. It is also for high-$e$ orbits that the assumption of a constant lag-angle (inherent to the constant-$Q$ model) is perhaps most suspect.

[Figure 9]

### 6.2 Excitation and Resonance Escape

We now return to libration excitation and the timing of escape for Mignard tides. Per Figures 6-8, during most of the Moon's outbound evolution in evection the libration amplitude is damped (*i.e.*, $\dot{\Theta}/\Theta < 0$) or undergoes only weak excitation. However as the Moon approaches the turn-around point in semi-major axis, there is a transition to increasing excitation. For low $A$, damped libration in the outbound phase rapidly transitions to excitation near the stall point (Figures 6 and 7a), reminiscent of the behavior seen in Touma and Wisdom (1998), suggesting that escape from resonance is likely to occur near this point, depending on the initial libration amplitude following capture. For larger values of $A$, libration amplitude growth past the stall point is more modest (Figure 7a), however, $\dot{\Theta}/\Theta$ remains positive throughout the Moon's subsequent orbital contraction, and its magnitude generally increases with time. This implies that resonant escape will occur well before the dual-synchronous end state is reached in the high-$A$ cases as well.

It is instructive to consider how cyclic variations in tidal strength lead to amplitude changes during a single libration cycle. First consider the lead constant term, $\mathcal{F}_s = 2\chi[\dot{\varepsilon}_T(\varepsilon_s) - \dot{\varepsilon}_s(\varepsilon_s)]$, in the forcing function from eqn. (6.5b). A constant $\dot{\varepsilon}_T(\varepsilon_s)$ during a state's counter-clockwise traverse of the upper level curve branch tries to push the trajectory across level curves. For specificity, on a level curve with turn-around points $\pm\Theta_o$ during the orbital contraction phase, $\dot{\varepsilon}_T(\varepsilon_s) < 0$, and the trajectory on the upper branch drifts downward toward level curves with more negative $\tilde{J}$ (see Figure 1f). As a result it encounters a turning point at $\Theta$ slightly less than $\Theta_o$. However, over the return, rightward trip on the lower branch, its continued downward motion causes the state to drift across level curves of higher $\tilde{J}$, reversing the process. The net result is a trajectory path that resembles a level curve, but whose point of symmetry is shifted off the $y$-axis to an angle $\sim \dot{\varepsilon}_T(\varepsilon_s)/40\chi\alpha\varepsilon_s < 0$ (see eqn. 5.8 for the case of a small offset angle).[5] This same path is repeated on future cycles unless there is a change in $\varepsilon_s$. A similar situation occurs for a constant $\dot{\varepsilon}_s(\varepsilon_s) < 0$ with $\dot{\varepsilon}_T = 0$, where a level curve with given turn-around points $\pm\Theta_o$ migrates down the y-axis.[6] This causes the state's position to be crossed by level curves with larger libration amplitudes, $\Theta > \Theta_o$ during its upper counter-clockwise traverse, but by curves of $\Theta < \Theta_o$ on its rightward lower return. In this case the trajectory again resembles the shape of a level curve shifted off the y-axis by $\sim -\dot{\varepsilon}_s(\varepsilon_s)/40\chi\alpha\varepsilon_s > 0$. Since during contraction the magnitude of $\dot{\varepsilon}_T$ is generally larger than $\dot{\varepsilon}_s$ (see Figure 5a), their combined influence yields the negative stationary state angle. But, for $\dot{\varepsilon}_T$ and $\dot{\varepsilon}_s$ that are constant during a libration cycle, there is no net change in libration amplitude.

---
[5] Recall that we define $\theta$ as the angle from the $y$-axis in the direction of the negative $x$-axis, so that on the upper branch of the level curve, $\dot{\theta} > 0$ corresponds to motion in the counter-clockwise direction. Because of this convention, a negative offset angle lies in quadrant I of our coordinate system, while a positive offset angle lies in quadrant II.

Now consider the second term in eqn. (6.5b), $\partial F/\partial \varepsilon|_{\varepsilon_s}(\varepsilon - \varepsilon_s) = 2\chi[\partial\dot{\varepsilon}_T/\partial\varepsilon(\varepsilon_s) - \partial\dot{\varepsilon}_s/\partial\varepsilon(\varepsilon_s)](\varepsilon - \varepsilon_s)$, describing cyclic variations of the tidal strengths. Figure 6a displays the partial derivatives $\partial\dot{\varepsilon}_T/\partial\varepsilon$, $\partial\dot{\varepsilon}_s/\partial\varepsilon$ during the evolution shown in Figure 4. Although during most of the orbit contraction phase the magnitude of $\dot{\varepsilon}_T$ exceeds that of $\dot{\varepsilon}_s$ (see Figure 5a), in Figure 6a we see that $|\partial\dot{\varepsilon}_s/\partial\varepsilon| > |\partial\dot{\varepsilon}_T/\partial\varepsilon|$, and that both derivatives are negative. Thus the quantity $[\partial\dot{\varepsilon}_T/\partial\varepsilon(\varepsilon_s) - \partial\dot{\varepsilon}_s/\partial\varepsilon(\varepsilon_s)]$ is positive during contraction, so that when $(\varepsilon - \varepsilon_s) > 0$, there is positive forcing, while when $(\varepsilon - \varepsilon_s) < 0$, the libration amplitude is damped. However, the two effects do not exactly compensate because $(\varepsilon - \varepsilon_s) > 0$ for proportionally more of the libration cycle, and consequently, a cycle finishes with a larger amplitude then when it started.

In the early outbound phase in evection, libration amplitude is typically damped. If there were no tidal change in $\varepsilon_s$ in this phase, eqn. (6.9) would reduce to $\dot{\Theta}/\Theta \to (1/2)\partial\dot{\varepsilon}_T/\partial\varepsilon$, and when $\partial\dot{\varepsilon}_T/\partial\varepsilon < 0$ (as implied by damping) the libration amplitude could be driven to a vanishing small quantity. However, if $\dot{\varepsilon}_s \neq 0$, there is a limit to this. The eccentricity of the upper libration path at the y-axis is $\varepsilon_s + \sqrt{\varepsilon_s^2 + 2\tilde{J}}$, implying a path half-width of $w = \sqrt{\varepsilon_s^2 - \varepsilon_\Theta^2}$, the final form employing eqn. (6.2a). Assuming $\Theta$ is small, $\cos 2\Theta \approx 1 - 2\Theta^2$ and eqn. (6.1) reads $\varepsilon_\Theta \approx \varepsilon_s - 10\alpha\Theta^2$. Accordingly, $w \approx \Theta\sqrt{20\alpha\varepsilon_s}$ to lowest order in $\Theta$. In a like manner to section 4.2, when the distance, $\sim \pi\dot{\varepsilon}_s/\omega$, the stationary point migrates over a half cycle is comparable to $w$, further decrease of $\Theta$ is thwarted by the evolving level curves pattern. This implies that the amplitude will not decrease below a characteristic value $\Theta_{min} \approx \pi|\dot{\varepsilon}_s/\varepsilon_s|/40\chi\alpha$. This value depends inversely on $\chi = \Omega_\oplus t_T$. Thus for slower tidal evolution (*i.e.*, larger tidal time constant $t_T$, smaller terrestrial $\Delta t$), the libration amplitude can be decreased to smaller values during the initial damped outbound phase.

Once excitation begins at time $\tau_{ex}$, integrating eqn. (6.9) gives

$$\ln\left(\frac{\Theta}{\Theta_{ex}}\right) = \frac{1}{2}\int_{\tau_{ex}}^{\tau}\left(\frac{\partial\dot{\varepsilon}_T}{\partial\varepsilon} - \frac{\partial\dot{\varepsilon}_s}{\partial\varepsilon}\right)d\tau \qquad (6.14)$$

where $\Theta_{ex} \approx \Theta_{min}(\tau_{ex})$ denotes the amplitude at $\tau_{ex}$. Thus the amplitude grows as[7]

$$\Theta(\tau) = \Theta_{min}(\tau_{ex})\exp\left[\frac{1}{2}\int_{\tau_{ex}}^{\tau}\left(\frac{\partial\dot{\varepsilon}_T}{\partial\varepsilon} - \frac{\partial\dot{\varepsilon}_s}{\partial\varepsilon}\right)d\tau\right] \qquad (6.15)$$

For a given evolution, one can integrate (6.15) to estimate when $\Theta \to \pi/2$ and escape occurs as a function of $A$ and the absolute rate of tidal evolution given by $\chi$. For $A \leq 10$, $L_0 = 2L_{EM}$, a non-synchronously rotating Moon, and $1\times 10^6 \leq t_T \leq 2\times 10^7$ (corresponding approximately to $80 < (Q/k_T) < 1800$), escape occurs early when the Earth-Moon system angular momentum has been reduced by only about 8% to 9% relative to its starting value. For $A = 10$, $L_0 = 2L_{EM}$, and a synchronously rotating moon (including the effects of permanent figure torques), the change in AM is even less, about 3% to 5%.

---

[6] In this case, there is a change in the Jacobi value associated with $\Theta_o$ found from eqn. (6.2b), *viz.*, $\tilde{J} = -(\varepsilon_* + 5\alpha\cos 2\Theta_o)^2/2$.

[7] We caution that this is an approximation since the form of $\dot{\Theta}/\Theta$ was derived for small amplitude.

## 7. Summary and Discussion

We have examined the tidal evolution of the Sun-Moon evection resonance employing the tidal model developed by Mignard (1980). This has been motivated by the work of Ćuk and Stewart (2012; CS12) who found a large decrease in the Earth-Moon system angular momentum (AM) due to this mechanism. Although the direct solar tidal torque on the Earth can drain its spin angular momentum, the loss is very small over the age of the solar system. In contrast, the evection resonance allows the Sun to indirectly drain the Earth's spin by exerting a torque on the lunar orbit that can then be transmitted to the Earth via the much stronger lunar tidal torque. Initial capture of the Moon into evection is not guaranteed. However, the case has been made that capture is probable given the slow outward tidal evolution rates associated with a fluid-like Earth in the aftermath of a Moon-forming giant impact (Zahnle *et al.*, 2015). If the evection resonance is then maintained, the loss of angular momentum could potentially be very large.

CS12 utilized a tidal model intended to approximate a constant-$Q$ model, in which the tide is assumed to form at a fixed angle ahead or behind the line connecting the centers of the tidally interacting objects. In order to avoid discontinuity at the synchronous orbit, they multiplied their tidal torque by a smoothing factor. A detailed analysis of the CS12 tidal model and its differences from a conventional constant-$Q$ model (Kaula, 1964) is contained in Wisdom and Tian (2015). They implemented a true constant-$Q$ model, and found that if the Moon's tidal parameters are assumed constant with time, the successful cases identified in CS12 remove too much angular momentum. Tian *et al*. (2017) subsequently demonstrated that tidal heating during the high-eccentricity evolution in evection invoked in CS12 would alter tidal dissipation in the Moon and cause rapid exit from formal resonance with little or no AM drain, again assuming a constant-$Q$ tidal model.

In this paper, we adopt the Mignard tidal model as utilized by Touma and Wisdom (1998) but apply it to an Earth-Moon system that initially has a much higher angular momentum than its current value, as considered in CS12. Our findings include:

(1) Angular momentum is drained from the Earth-Moon system while the evection resonance is maintained. In the limiting case of a moon that remained continually in resonance, evection would drive the system to a co-synchronous end state, $s = s_m = n$, with a final angular momentum independent of the system's initial angular momentum. For the Earth-Moon system this limiting state was never reached, implying that either the Moon was never captured into evection, or that it escaped from resonance. In the latter case, the timing of escape determines the degree of angular momentum modification due to formal evection, with increased angular momentum drain as the time spent in resonance lengthens.

(2) During resonance, the Moon's longitude of perigee librates about the stationary state angle (which is approximately ± 90° from the Earth-Sun line). Escape from resonance requires the resonant trajectory to cross the separatrix boundary, which can occur if the libration amplitude, Θ, approaches $\pi/2$. Tidal evolution causes libration excitation and/or damping if there is a variation in tidal strength over a libration cycle.

(3) For Mignard tides, resonant libration is damped or minimally excited during most of the Moon's initial outward expansion in resonance. However, as the Moon approaches the "stall" point (after which its orbit contracts), libration amplitude excitation increases and remains positive throughout the rest of the evolution. This is true across a wide range of tidal parameters and for either a non-synchronously rotating moon with no permanent figure torques, or a synchronously rotating, triaxial moon.

(4) We estimate that libration excitation leads to escape from resonance early in the evolution, resulting in ≤ 10% angular momentum loss for an Earth-Moon system with an initial AM that is roughly twice that of the current Earth-Moon. This is similar to early resonance escape seen in Touma and Wisdom (1998) with Mignard tides for lower AM systems.

We conclude that with Mignard tides, formal evection resonance does not appear capable of reconciling high-angular momentum giant impact models (Ćuk and Stewart, 2012; Canup, 2012) with the current Earth-Moon system. This result augments those of Wisdom and Tian (2015) and Tian *et al*. (2017), who conclude that formal evection is unsuccessful in reproducing the Earth-Moon AM for constant-$Q$ tides.

Alternatively, appropriate angular momentum removal to accommodate a high-AM Moon-forming giant impact could result from effects other than formal libration in evection. With constant-$Q$ tides, Wisdom and Tian (2015) identified an evection-related limit cycle in which large amounts of AM can be extracted from the Earth-Moon even though the Moon is not librating within resonance; a broadly similar "quasi-resonance" was seen in preliminary integrations using the Mignard model by Ward and Canup (2013) and Rufu and Canup (2019). Such effects are not accessible with the methods here. It has also been proposed (Ćuk *et al*., 2016) that an entirely different mechanism could have reduced the early Earth-Moon AM, involving an initial Earth with a very high obliquity and a Laplace plane instability as the lunar orbit expands. However, the range of successful parameters for this mechanism remains unclear.

The analytic developments here include simplifications, notably coplanar dynamics, an evolution description limited to 4$^{th}$ order in eccentricity, and an assumption of small libration amplitude when assessing how the amplitude varies with time. Ultimately, integration of the system's full evolution in $a, s, s_m, e$, and $\theta$ is needed to assess the behavior of evection in the context of the Mignard tidal model, which will be a topic of a subsequent paper. Additional effects not considered here include the potential time-dependence of the tidal parameters during evolution in evection, and the potential for spin-orbit resonances in the Moon's rotation state that differ from the non-synchronous or synchronous rotations considered here.

**Acknowledgements**

This research was supported by NASA's SSERVI and Emerging Worlds programs. RR is an Awardee of the Weizmann Institute of Science - National Postdoctoral Award Program for Advancing Women in Science. We thank the two reviewers for their constructive comments that greatly improved the paper, and Jack Wisdom for helpful discussion. The equations provided in the main text provide all information needed to reproduce the results presented here.

**Table 1: Some Variable Definitions**

| | |
|---|---:|
| Semi-major axis, mean motion, and eccentricity of Moon | $a, n, e$ |
| Earth mass and radius | $M, R$ |
| Lunar mass and radius | $m, R_m$ |
| Mass ratio $m/M$ | $\mu$ |
| Earth spin rate, lunar spin rate | $s, s_m$ |
| Angular momentum of Earth-Moon system | $L_{EM}$ |
| Angular momentum of lunar orbit | $L_{orb}$ |
| Measure of the strength ratio of lunar to Earth tides | $A$ |
| Maximum principal moments of inertia of Earth, Moon | $C, C_m$ |
| Ratio of principal moments $C_m/C$ | $\kappa$ |
| Gyration constant for Earth | $\lambda$ |
| Ratio of $\mu/\lambda$ | $\gamma$ |
| Circumterrestrial orbital frequency at $R$ | $\Omega_\oplus$ |
| Earth's orbital frequency about the Sun | $\Omega_\odot$ |
| Tidal lag times, lag angle, and Love numbers for Earth, Moon | $\Delta t, \Delta t_m, \delta, k_T, k_m$ |
| Tidal evolution time constant | $t_T$ |
| Normalized tidal evolution time | $\chi = \Omega_\odot t_T$ |
| Libration amplitude | $\Theta$ |

**Table 2: Tidal Polynomials and their Derivatives**

$$\tilde{f}_1(\varepsilon) = 1 + \frac{15}{2}\varepsilon + \frac{45}{8}\varepsilon^2 + \frac{5}{16}\varepsilon^3 \quad ; \quad \frac{\partial \tilde{f}_1(\varepsilon)}{\partial \varepsilon} = \frac{15}{2} + \frac{45}{4}\varepsilon + \frac{15}{16}\varepsilon^2$$

$$\tilde{f}_2(\varepsilon) = 1 + \frac{31}{2}\varepsilon + \frac{255}{8}\varepsilon^2 + \frac{185}{16}\varepsilon^3 + \frac{25}{64}\varepsilon^4 \quad ; \quad \frac{\partial \tilde{f}_2(\varepsilon)}{\partial \varepsilon} = \frac{31}{2} + \frac{255}{4}\varepsilon + \frac{555}{16}\varepsilon^2 + \frac{25}{16}\varepsilon^3$$

$$\tilde{g}_1(\varepsilon) = \frac{11}{2} + \frac{33}{4}\varepsilon + \frac{11}{16}\varepsilon^2 \quad ; \quad \frac{\partial \tilde{g}_1(\varepsilon)}{\partial \varepsilon} = \frac{33}{4} + \frac{11}{8}\varepsilon$$

$$\tilde{g}_2(\varepsilon) = 9 + \frac{135}{4}\varepsilon + \frac{135}{8}\varepsilon^2 + \frac{45}{64}\varepsilon^3 \quad ; \quad \frac{\partial \tilde{g}_2(\varepsilon)}{\partial \varepsilon} = \frac{135}{4} + \frac{135}{4}\varepsilon + \frac{135}{64}\varepsilon^2$$

**Figure Captions**

Fig. 1. Level curves for the evection resonance for different energies. The Sun is in the direction of the positive $x$-axis. (a) $\varepsilon_*/5\alpha = -4$: Pre-resonance where all motion is counter-clockwise circulation about the origin. (b) $\varepsilon_*/5\alpha = -1$: First appearance of stable stationary states on $y$-axis. (c) $\varepsilon_*/5\alpha = 0$: Shallow resonance where the level curve $\tilde{J} = 0$ is a separatrix dividing counter-clockwise libration about the stationary point from level curves circulating the origin. (d) $\varepsilon_*/5\alpha = 1$: First appearance of unstable saddle points on the $x$-axis. (e) $\varepsilon_*/5\alpha = 2$: Deep resonance with a separatrix composed of two branches emanating from the saddle points. Below the lower branch are level curves circulating the origin in the clockwise direction. (f) $\varepsilon_*/5\alpha = 4$: Still further into deep resonance, with the saddle points farther apart.

Fig. 2. Schematic of level curve domains in deep resonance (location of outer $\Upsilon_1$ domain not shown to scale).

Fig. 3. (a) The initial system angular momentum, $L_o$, and starting Earth spin, $s'_o$, (assuming an initial near circular orbit of the Moon at three Earth radii) that would result in evection resonance location $a_{\text{res}}$. Also shown (dashed curves) are the Earth spin, $s'$, and lunar orbit angular momentum, $L_{\text{orb}}$, at resonance encounter. The current Earth-Moon angular momentum, $L_{EM}$, as well as twice its value are shown for comparison (dotted lines).

Fig. 4. Tidal evolution of the Earth-Moon system in evection with damped libration for $A = 10$ and starting angular momentum $L_o = 2L_{EM}$. (a) Scaled lunar semi-major axis, $a'$, as a function of time. (b) The stationary state eccentricity squared, $\varepsilon_s \equiv e_s^2$, vs. time. (c) The stationary state eccentricity squared vs. lunar semi-major axis during the evolution. (d) Time variation of the system angular momentum, $L$, the Earth and Moon spin rates, $s, s_m$, the angular momentum of the lunar orbit, $L_{\text{orb}}$ and its mean motion, $n$. The gray area represents the stage where eccentricity is increasing.

Fig. 5. (a) Time variations of the semimajor-axis derivative, $\dot{a}'/a'$, stable eccentricity, $\dot{\varepsilon}_s$, and eccentricity derivatives due to lunar and Earth tides, $\dot{\varepsilon}_T$; The gray area represents the stage where eccentricity is increasing, $\dot{\varepsilon}_s > 0$. (b) Time variations of the orbital AM, $\dot{L}'_{\text{orb}}$, Earth's spin, $\dot{s}'$ and total AM, $\dot{L}'$, for the evolution in Figure 4. Before resonance capture, the increase in orbital AM is compensated by the decrease in the planet's spin, hence the total AM is constant ($\dot{L}' = 0$). During the outward phase after the resonance capture, the total AM decreases as both the orbital AM and spin rate decrease. During the inward migration stage, $\dot{\varepsilon}_s < 0$, the orbital AM remains relatively constant, while the total AM decreases due to the slowdown of Earth's spin, $\dot{L}' \sim \dot{s}'$.

Fig. 6. (a) Partial derivatives of $\partial \dot{\varepsilon}_s/\partial \varepsilon$ (solid red) and $\partial \dot{\varepsilon}_T/\partial \varepsilon$ (dashed dark red) of the evolution depicted in Figure 4. (b) Rate of change for the libration amplitude, $\dot{\Theta}/\Theta$. During most of the outward migration (gray area) the libration amplitude decreases, maintaining formal resonance. Near the turnaround point, the libration amplitude increases, promoting resonance escape.

Fig. 7. The evolution of the rate of change of libration amplitude ($\dot{\Theta}/\Theta$) for (a) Varied $A$ values for $L_0 = 2L_{EM}$ and (b) Varied $L_0/L_{EM}$ values for $A = 10$. With larger $A$ values, the transition

between the libration amplitude damping and excitation is more gradual. With larger initial AM values, the libration amplitude damping stage, which promotes resonance occupancy, is longer.

Fig. 8. Evolution of the rate of change of libration amplitude ($\dot{\Theta}/\Theta$) for synchronous rotation maintained by a permanent figure torque with $L_o = 2L_{EM}$ and $A = 10$ (gray line), with non-synchronous rotation case shown for comparison (black line). The libration amplitude excitation for the synchronous rotation case is more gradual compared to the non-synchronous case, hence the resonance escape is delayed. Note that the AM removal rate in the synchronous case is lower than the non-synchronous case (see Fig. A3), hence despite this delay, the overall amount of AM removed by evection is reduced.

Fig. 9. Evolution of the rate of change of libration amplitude ($\dot{\Theta}/\Theta$) using the constant-$Q$ tidal model for a synchronously rotating Moon, with tidal expressions and associated tidal $A$ constant defined by eqns. (12) and (21) to (40) in Wisdom and Tian (2015), for $L_o = 2L_{EM}$, $Q_\oplus = 400$, and varied $A$ values. With a constant-$Q$ model and $A = 1.7$ and 2, eqn. (6.9) predicts an extended period of libration amplitude damping ($\dot{\Theta}/\Theta < 0$) even as the Moon's semi-major axis contracts (orbit contraction for these cases commences at $t \leq 15$ [$10^4$ yr]). This implies protracted resonance occupancy, consistent with simulations of Wisdom and Tian for this narrow range of $A$ values (*e.g.*, their Figs. 2 and 9). In contrast, for $A \geq 3$ (darker blue lines) increasingly strong amplitude excitation is predicted, suggesting limited resonance occupancy. Wisdom and Tian found minimal or no libration in formal evection for constant-$Q$ tides and these larger $A$ values.

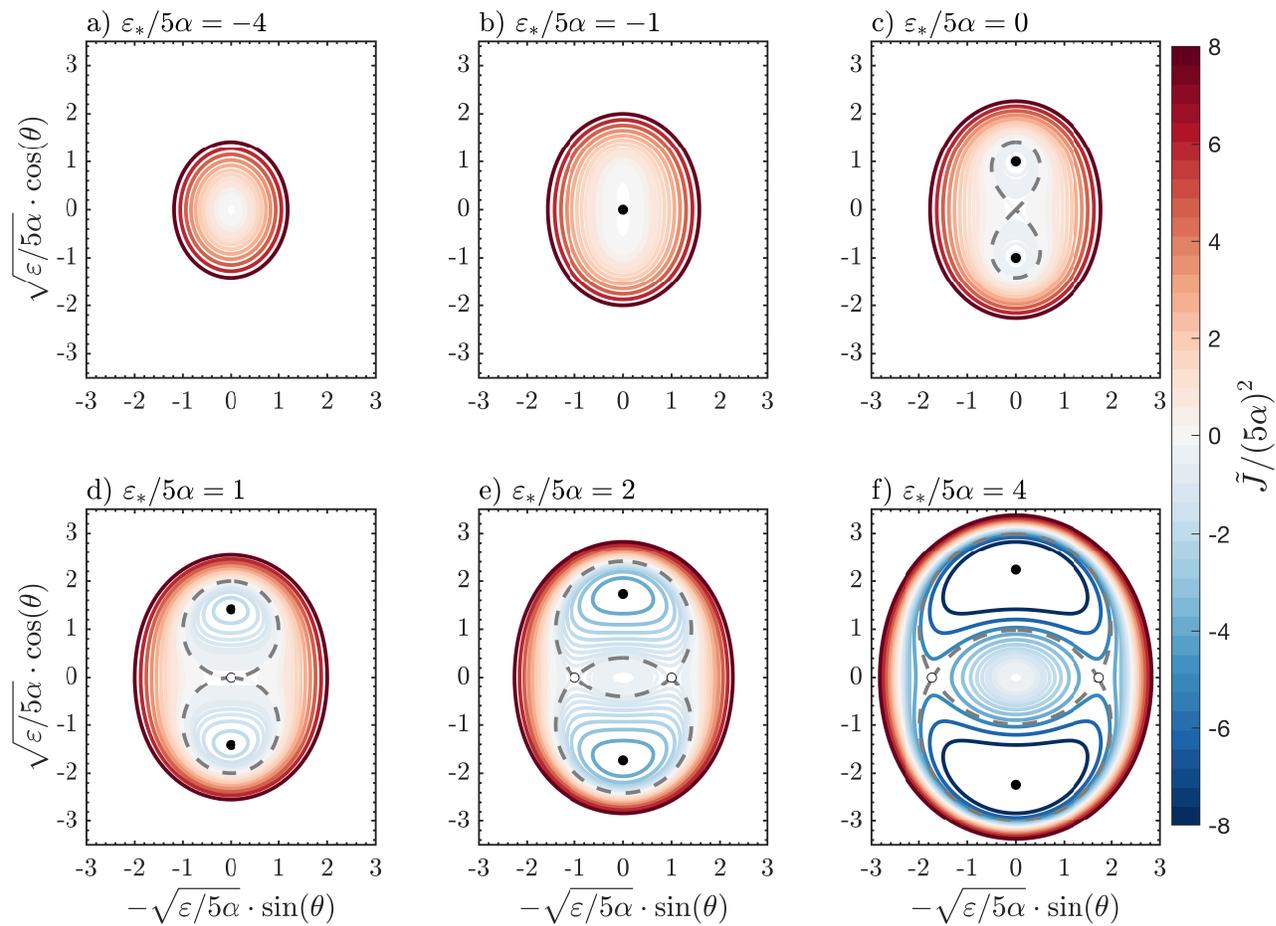

Figure 1

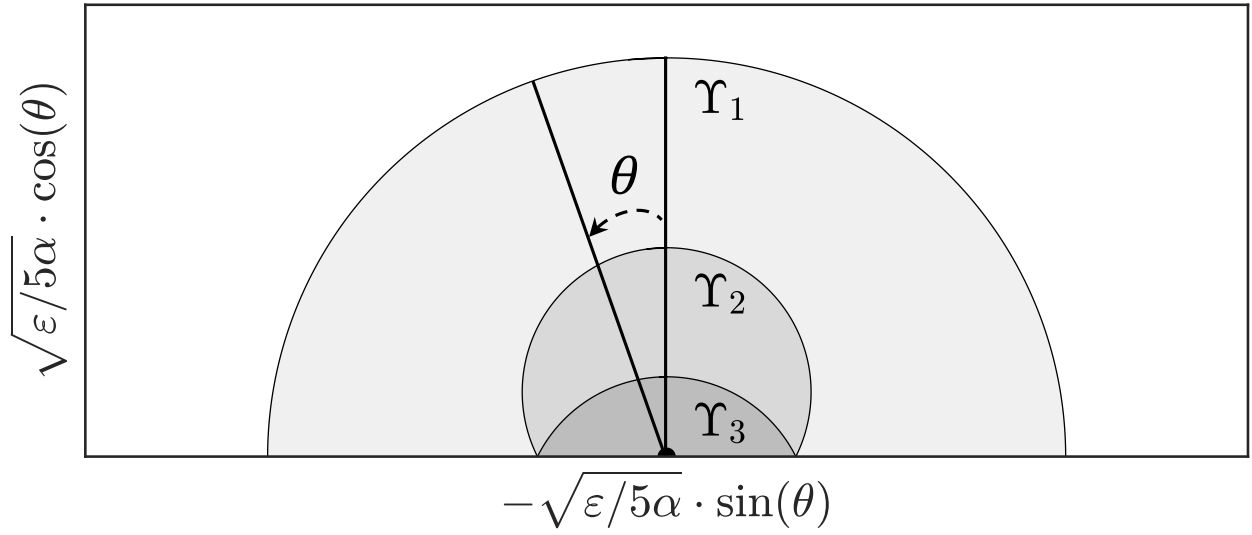

Figure 2

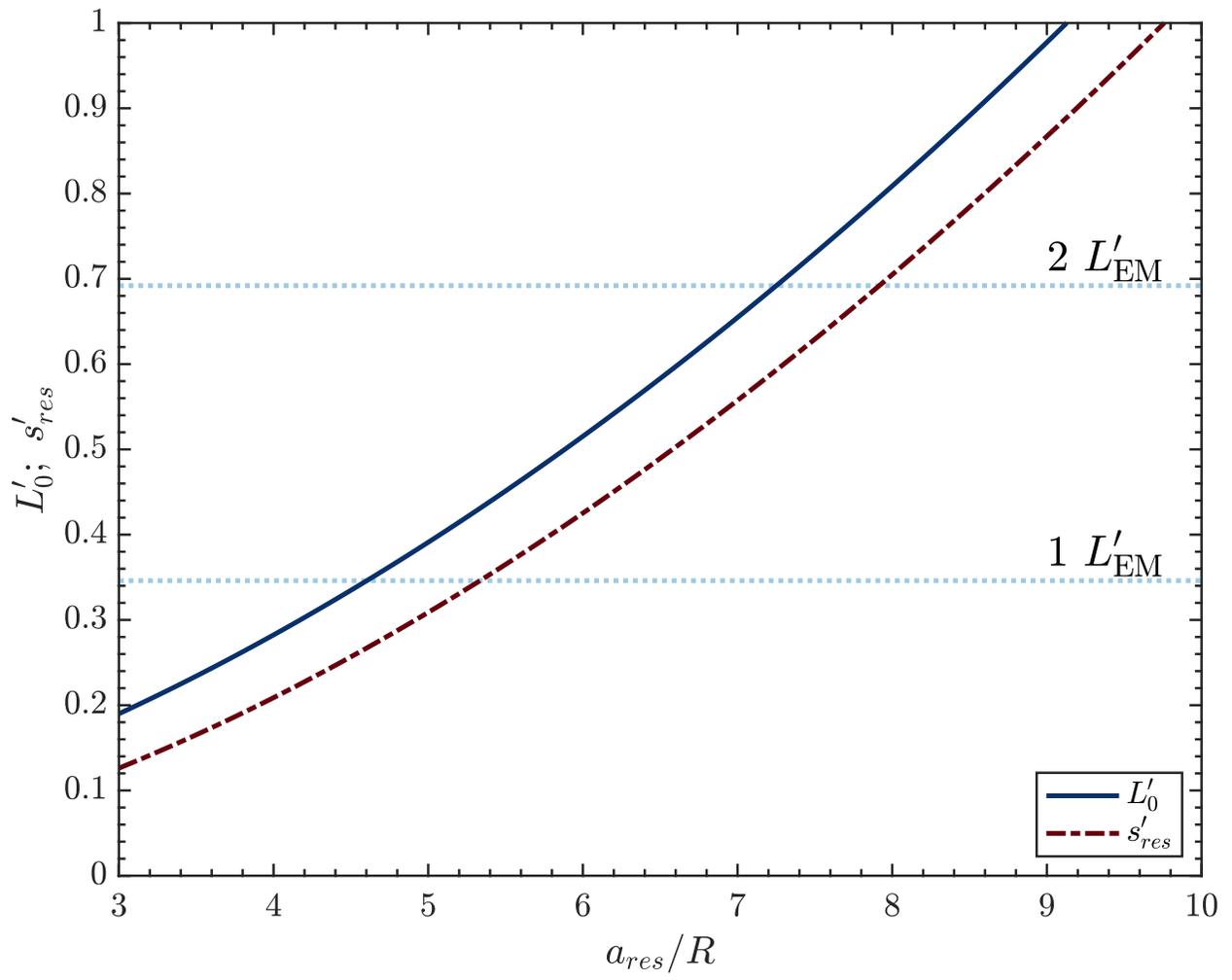

Figure 3

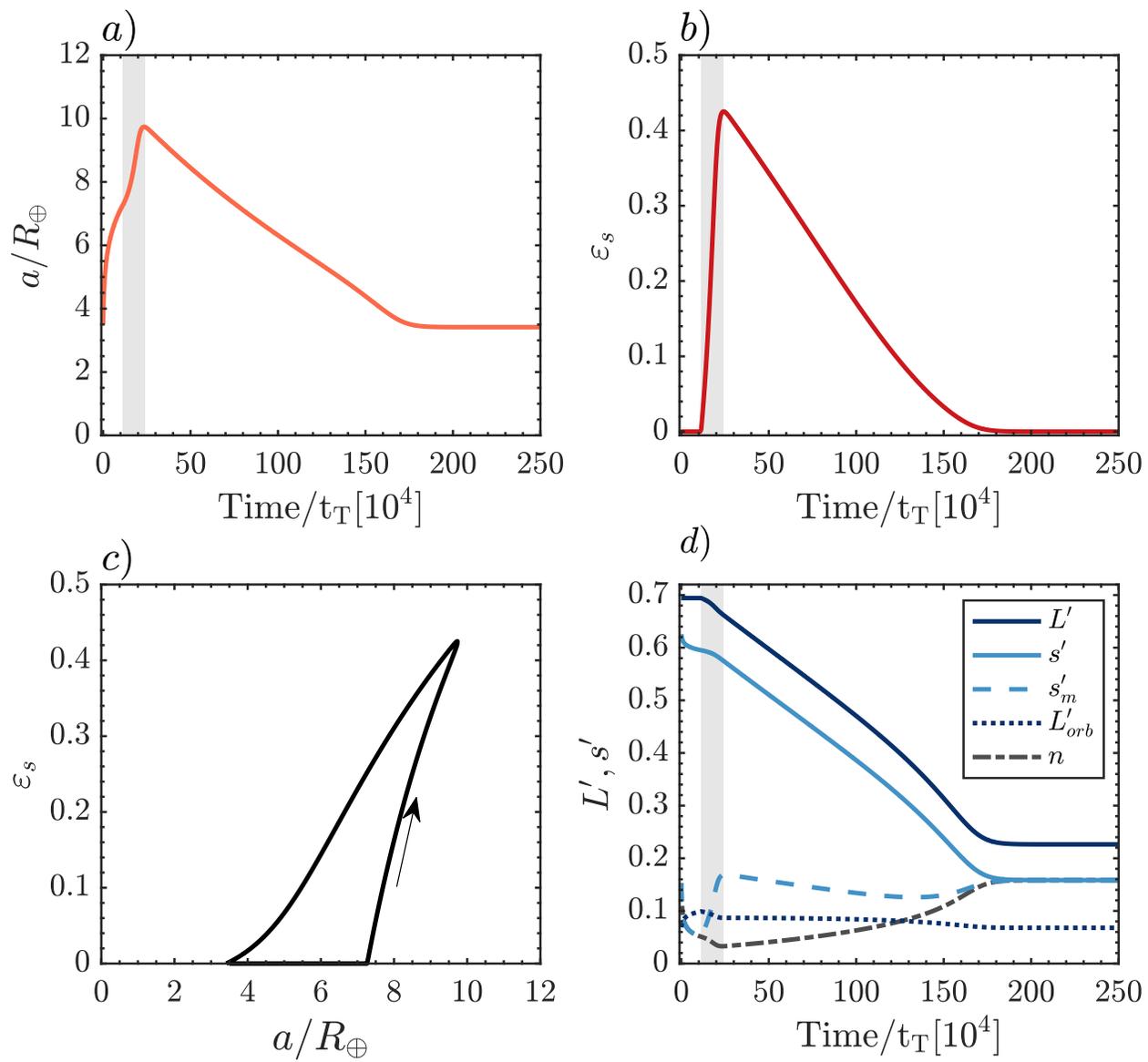

Figure 4

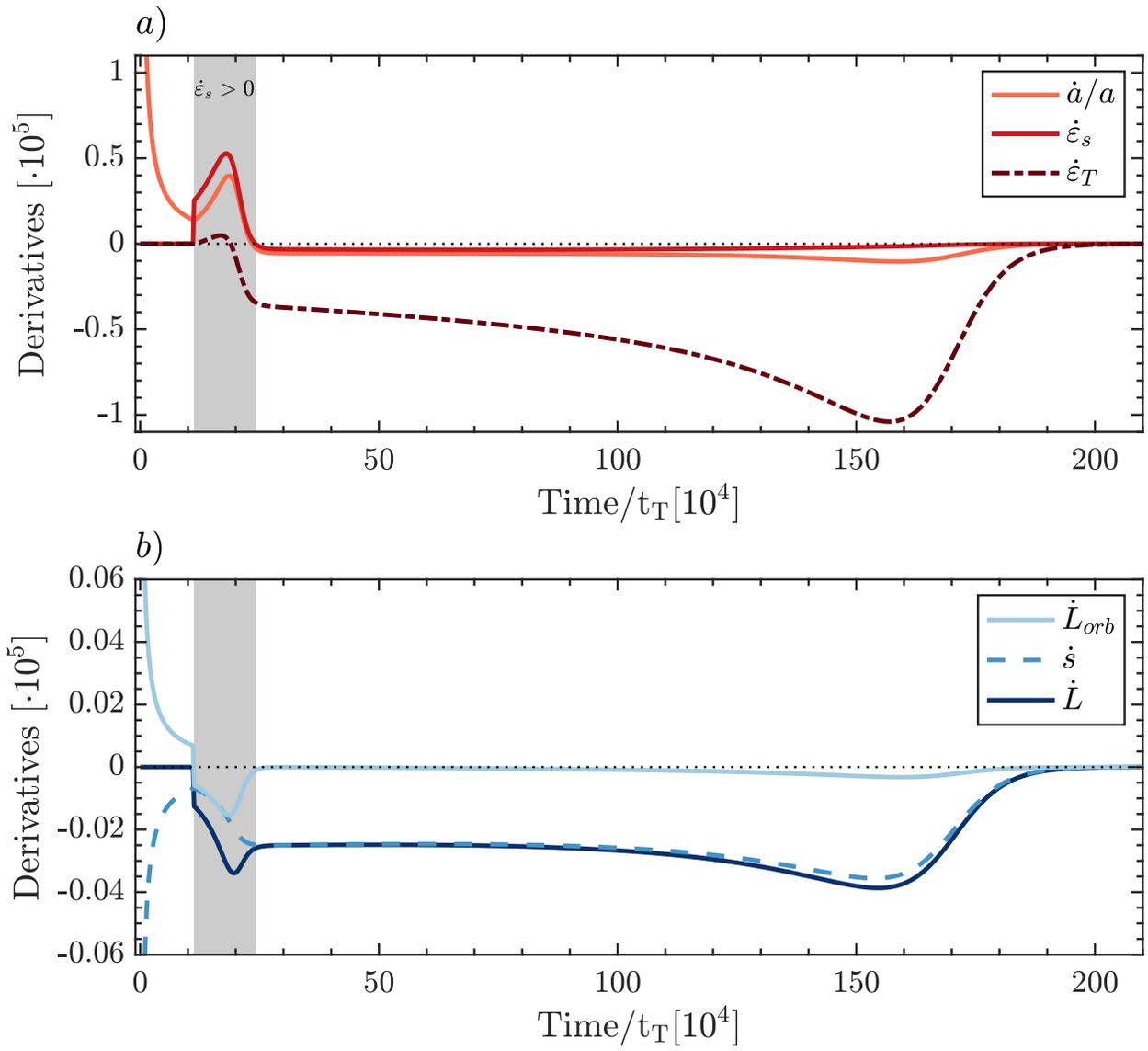

Figure 5

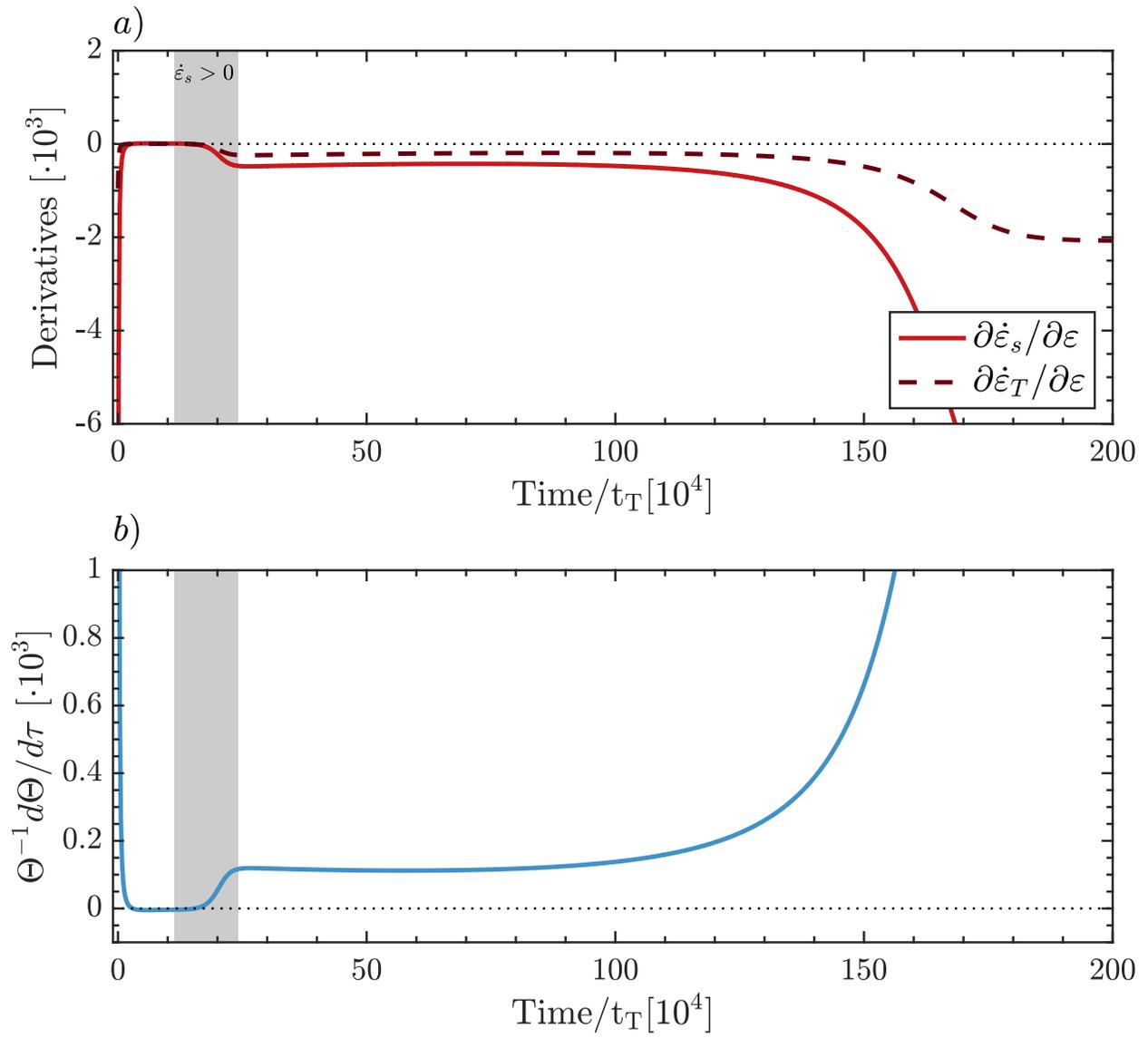

Figure 6

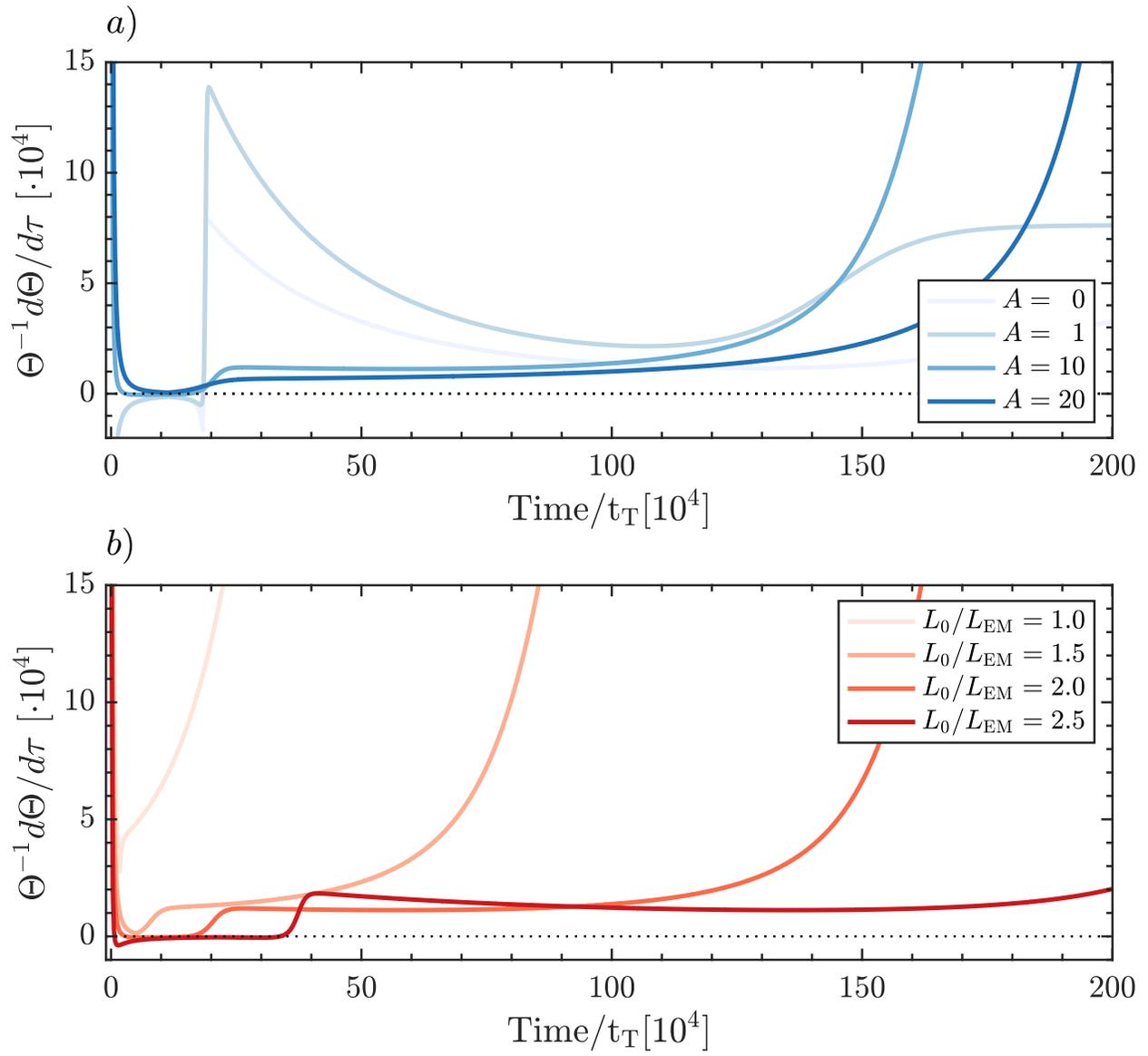

Figure 7

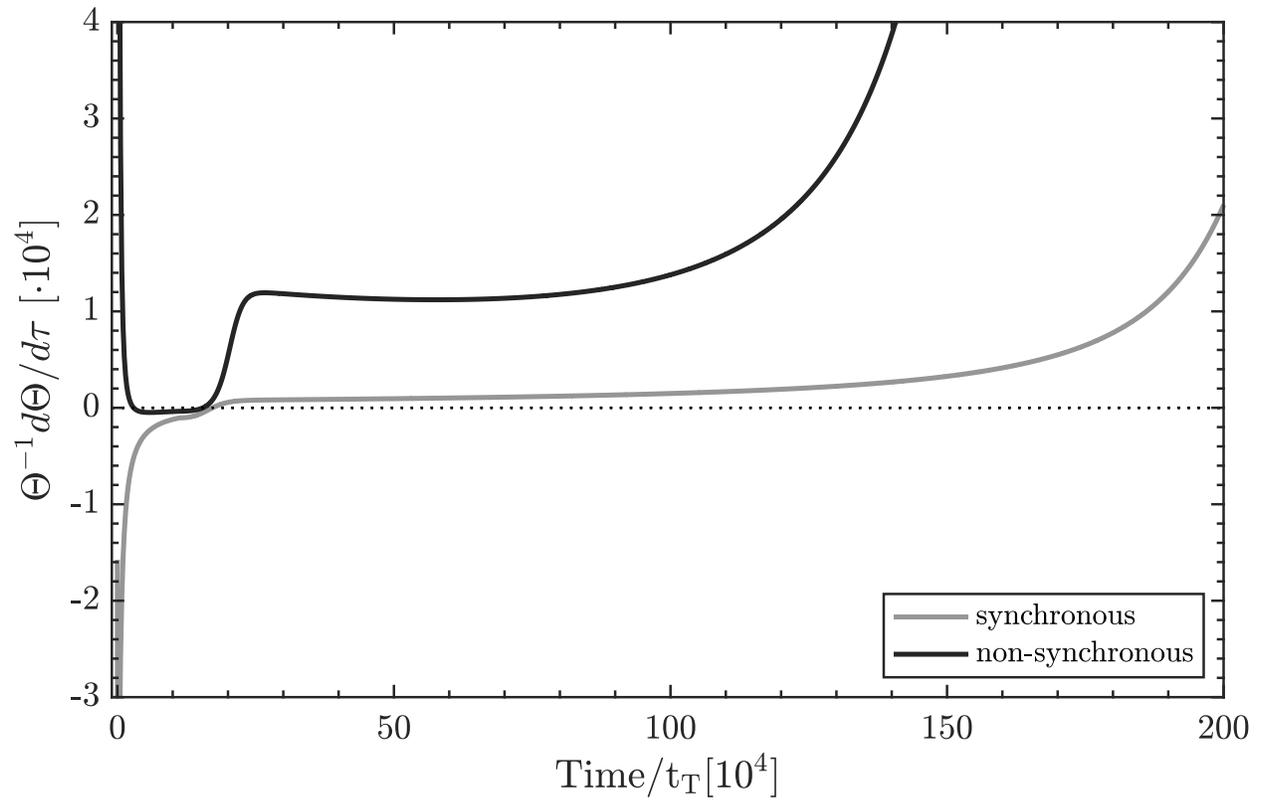

Figure 8

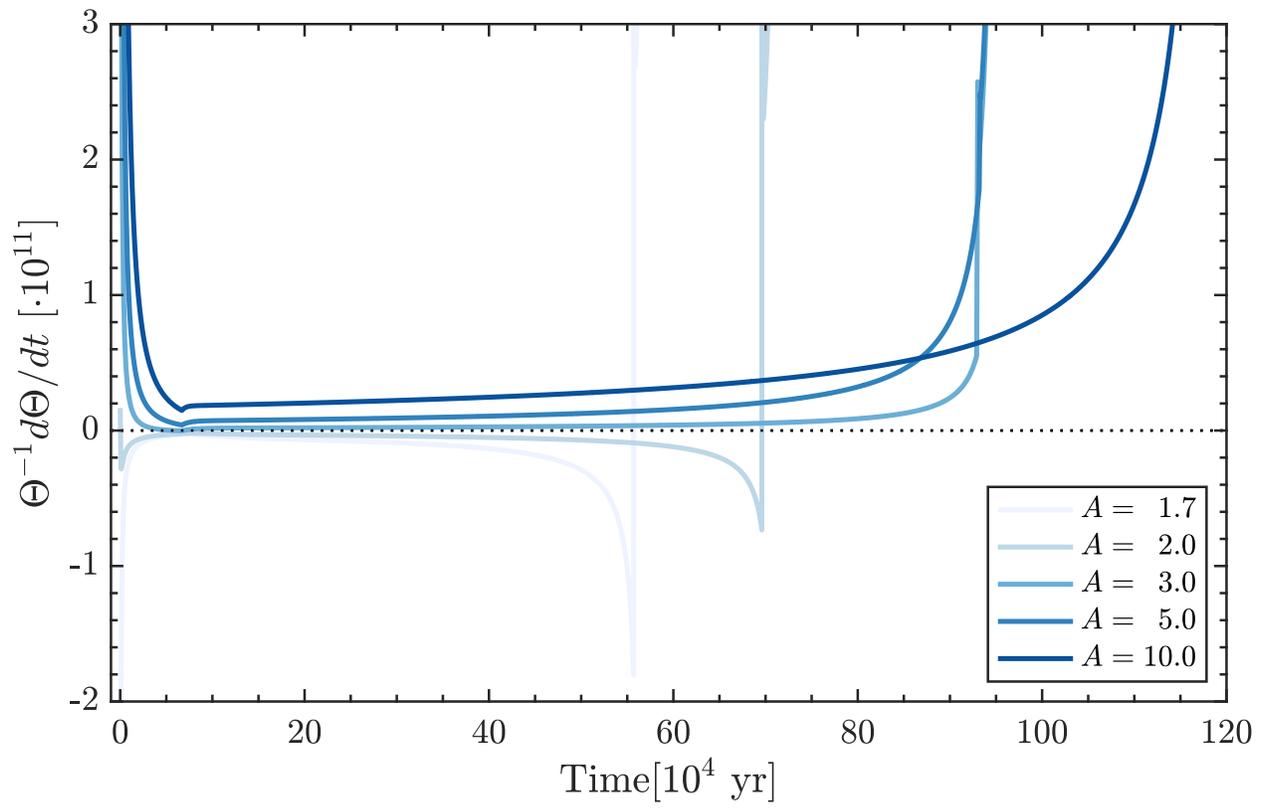

Figure 9

Supporting Information for

**Analytical Model for the Tidal Evolution of the Evection Resonance and the Timing of Resonance Escape**

William R. Ward, Robin M. Canup, and Raluca Rufu

Planetary Science Directorate Southwest Research Institute; Boulder, CO 80302

**Contents of this file**

Appendix A to E

Figures A1 to A3

**Appendix A - Jacobi Constant**

The lunar equation of motion with Earth's oblateness and Sun's influence treated as a disturbing potential, $\mathcal{R} = -(\Phi_\oplus + \Phi_\odot)$, in the coordinate system centered on Earth is:

$$\ddot{\vec{r}} = \nabla(U + \mathcal{R}) \tag{A1}$$

where $U \approx GM/r$ is the two-body Earth-Moon potential. Here the Moon's mass is ignored, and the lunar inclination and terrestrial obliquity are assumed negligible ($\ddot{z} = 0$), hence:

$$\ddot{x} = \frac{\partial}{\partial x}(U + \mathcal{R}) = GM \frac{\partial}{\partial x}\left(\frac{1}{\sqrt{x^2+y^2}}\right) + \frac{\partial \mathcal{R}}{\partial x} \tag{A2a,b}$$

$$\ddot{y} = \frac{\partial}{\partial y}(U + \mathcal{R}) = GM \frac{\partial}{\partial y}\left(\frac{1}{\sqrt{x^2+y^2}}\right) + \frac{\partial \mathcal{R}}{\partial y}$$

This can be rearranged to (Brouwer and Clemence, 1961):

$$\ddot{x} + \frac{GMx}{r^3} = \frac{\partial \mathcal{R}}{\partial x} \tag{A3a, b}$$

$$\ddot{y} + \frac{GMy}{r^3} = \frac{\partial \mathcal{R}}{\partial y}$$

We switch to a rotating coordinate system $(\mathcal{X}, \mathcal{Y})$, where the $\mathcal{X}$ axis is aligned along the Earth-Sun line ($\vec{r}'$) and the system rotates with an angular velocity of $\Omega_\odot$ (assuming that Earth's orbit around the Sun is circular), with (Murray and Dermott, 1999):

$$x = \mathcal{X} \cos \Omega_\odot t - \mathcal{Y} \sin \Omega_\odot t \qquad \text{(A4a,b,c,d)}$$

$$y = \mathcal{X} \sin \Omega_\odot t + \mathcal{Y} \cos \Omega_\odot t$$

$$\ddot{x} = (\ddot{\mathcal{X}} - 2\Omega_\odot \dot{\mathcal{Y}} - \Omega_\odot^2 \mathcal{X}) \cos \Omega_\odot t - (\ddot{\mathcal{Y}} + 2\Omega_\odot \dot{\mathcal{X}} - \Omega_\odot^2 \mathcal{Y}) \sin \Omega_\odot t$$

$$\ddot{y} = (\ddot{\mathcal{X}} - 2\Omega_\odot \dot{\mathcal{Y}} - \Omega_\odot^2 \mathcal{X}) \sin \Omega_\odot t + (\ddot{\mathcal{Y}} + 2\Omega_\odot \dot{\mathcal{X}} - \Omega_\odot^2 \mathcal{Y}) \cos \Omega_\odot t$$

Substituting these relations into the equations of motion, multiplying (A.3a) by $\cos \Omega_\odot t$, and (A.3b) by $\sin \Omega_\odot t$, and adding the results gives

$$\ddot{\mathcal{X}} - 2\Omega_\odot \dot{\mathcal{Y}} - \Omega_\odot^2 \mathcal{X} + \frac{GM\mathcal{X}}{r^3} = \frac{\partial \mathcal{R}}{\partial \mathcal{X}} \qquad \text{(A5)}$$

where the LHS of the equation is given by the chain rule: $\frac{\partial \mathcal{R}}{\partial \mathcal{X}} = \frac{\partial \mathcal{R}}{\partial x}\frac{\partial x}{\partial \mathcal{X}} + \frac{\partial \mathcal{R}}{\partial y}\frac{\partial y}{\partial \mathcal{X}} = \frac{\partial \mathcal{R}}{\partial x} \cos \Omega_\odot t + \frac{\partial \mathcal{R}}{\partial y} \sin \Omega_\odot t$. Similarly, multiplying (A.3a) by $-\sin \Omega_\odot t$, and (A.3b) by $\cos \Omega_\odot t$, and adding the results:

$$\ddot{\mathcal{Y}} + 2\Omega_\odot \dot{\mathcal{X}} - \Omega_\odot^2 \mathcal{Y} + \frac{GM}{r^3} = \frac{\partial \mathcal{R}}{\partial \mathcal{Y}} \qquad \text{(A6)}$$

where the LHS of the equation is given by the chain rule: $\frac{\partial \mathcal{R}}{\partial \mathcal{Y}} = \frac{\partial \mathcal{R}}{\partial x}\frac{\partial x}{\partial \mathcal{Y}} + \frac{\partial \mathcal{R}}{\partial y}\frac{\partial y}{\partial \mathcal{Y}} = -\frac{\partial \mathcal{R}}{\partial x} \sin \Omega_\odot t + \frac{\partial \mathcal{R}}{\partial y} \cos \Omega_\odot t$. The last two expressions can be simplified by (Brouwer and Clemence, 1961):

$$\ddot{\mathcal{X}} - 2\Omega_\odot \dot{\mathcal{Y}} = \frac{\partial F}{\partial \mathcal{X}} \qquad \text{(A7a, b)}$$

$$\ddot{\mathcal{Y}} + 2\Omega_\odot \dot{\mathcal{X}} = \frac{\partial F}{\partial \mathcal{Y}}$$

where $F \equiv \frac{GM}{r} + \frac{\Omega^2}{2}(\mathcal{X}^2 + \mathcal{Y}^2) + \mathcal{R}$.

To get the Jacobi integral, we multiply (A.7a) by $\dot{\mathcal{X}}$, (A.7b) by $\dot{\mathcal{Y}}$ and add them:

$$\ddot{\mathcal{X}}\dot{\mathcal{X}} + \ddot{\mathcal{Y}}\dot{\mathcal{Y}} = \frac{\partial F}{\partial x}\dot{\mathcal{X}} + \frac{\partial F}{\partial y}\dot{\mathcal{Y}} \tag{A8}$$

Integrating the last expression:

$$\frac{1}{2}\dot{\mathcal{X}}^2 + \frac{1}{2}\dot{\mathcal{Y}}^2 = F + \frac{J}{m} \tag{A9}$$

where $J$ is the modified Jacobi constant (in energy units).

$$\frac{1}{2}(\dot{\mathcal{X}}^2 + \dot{\mathcal{Y}}^2) - \frac{GM}{r} - \frac{\Omega_\odot^2}{2}(\mathcal{X}^2 + \mathcal{Y}^2) - \mathcal{R} = \frac{J}{m} \tag{A10}$$

We return to the non-rotating frame, centered on Earth, to express the Jacobi constant in terms of the Moon's $a$ and $e$. We use the relation:

$$\dot{\mathcal{X}}^2 + \dot{\mathcal{Y}}^2 = \dot{x}^2 + \dot{y}^2 + \Omega_\odot^2(x^2 + y^2) + 2\Omega_\odot(\dot{x}y - \dot{y}x) \tag{A11}$$

(Note that $\mathcal{X}^2 + \mathcal{Y}^2 = x^2 + y^2$, since distances are invariant under rotation transformations) to yield

$$\frac{1}{2}(\dot{x}^2 + \dot{y}^2) + \Omega_\odot(\dot{x}y - \dot{y}x) - \frac{GM}{r} - \mathcal{R} = \frac{J}{m} \tag{A12}$$

The kinetic energy can be replaced by (Murray and Dermott, 1999),

$$\frac{1}{2}(\dot{x}^2 + \dot{y}^2) = GM\left(\frac{1}{r} - \frac{1}{2a}\right) \tag{A13}$$

and we set $\dot{y}x - \dot{x}y = \vec{r} \cdot \vec{v} = L_{orb}/m = \sqrt{GMa(1-e^2)}$. Substituting these into (A12) gives eqn. (2.5) in the main text,

$$J = m\left[-\frac{GM}{2a} - \mathcal{R} - \Omega_\odot\sqrt{GMa(1-e^2)}\right].$$

**Appendix B - Stationary States**

*Tidal free states*

Denoting $\tilde{\alpha} \equiv 2\alpha(1 - 5\cos 2\theta)$, eqn. (2.15) is rearranged as, $\varepsilon = 1 - \eta[1 - \tilde{\alpha}(1 - \varepsilon)^{1/2}]^{-1/2}$. In the limit $\tilde{\alpha} \to 0$, $\varepsilon \to 1 - \eta$. For small $\tilde{\alpha}$, we write $\varepsilon = 1 - \eta + \Delta\varepsilon$, to find

$$\Delta\varepsilon = \eta\{1 - [1 - \tilde{\alpha}(\eta - \Delta\varepsilon)^{1/2}]^{-1/2}\} \tag{B1}$$

To second order accuracy in $\alpha$, the RHS is explicitly expanded to second order in $\tilde{\alpha}$,

$$\Delta\varepsilon \approx -\eta\{\tfrac{1}{2}\tilde{\alpha}(\eta - \Delta\varepsilon)^{1/2} + \tfrac{3}{8}\tilde{\alpha}^2(\eta - \Delta\varepsilon)\}. \tag{B2}$$

Assuming $\Delta\varepsilon \sim \mathcal{O}(\alpha)$ as well, $(\eta - \Delta\varepsilon)^{1/2} \approx \eta^{1/2}(1 - \Delta\varepsilon/2\eta)$. This leads to

$$\Delta\varepsilon \approx -\tfrac{1}{2}\tilde{\alpha}\eta^{3/2}(1 + \tfrac{3}{4}\tilde{\alpha}\eta^{1/2})/(1 - \tfrac{1}{4}\tilde{\alpha}\eta^{1/2}) \approx -\tfrac{1}{2}\tilde{\alpha}\eta^{3/2}(1 + \tilde{\alpha}\eta^{1/2}), \tag{B3}$$

and accordingly, $\varepsilon \approx 1 - \eta - \tilde{\alpha}\eta^{3/2}/2 - \tilde{\alpha}^2\eta^2/2$. To lowest order in $\alpha$, this reduces to $\varepsilon \approx 1 - \eta - \alpha(1 - 5\cos 2\theta)\eta^{3/2}$. Solution of this equation at $\theta = 0, \pi$ yields the y-axis stationary point value, $\varepsilon_s \approx 1 - \eta + 4\alpha\eta^{3/2}$, while the unstable stationary points at $\theta = \pm \pi/2$ on the x-axis are located at $\varepsilon_{sx} \approx 1 - \eta - 6\alpha\eta^{3/2}$. Their average value is $\varepsilon_* \approx 1 - \eta - \alpha\eta^{3/2}$.

We further simply by neglecting terms of order $\alpha\varepsilon$; combining the above expression for $\varepsilon_*$ with $\eta \approx (1 - \varepsilon)$ from eqn. (2.15) then gives $\varepsilon_* \approx 1 - \eta - \alpha(1 - \varepsilon)^{3/2} \approx 1 - \eta - \alpha$, $\varepsilon_s \approx 1 - \eta + 4\alpha = \varepsilon_* + 5\alpha$, and $\varepsilon_{sx} \approx 1 - \eta - 6\alpha = \varepsilon_* - 5\alpha$.

*Tidal states*

Because tides displace the stationary angle off the y-axis, there is a net average solar torque. The solar torque at the stationary point from eqn. (3.2b) becomes

$$T' = 10\gamma\chi a'^{1/2}\alpha\varepsilon_s \sin 2\theta_s = (\gamma/2)a'^{1/2}(\dot{\varepsilon}_T - \dot{\varepsilon}), \tag{B4}$$

while the rate at which the system angular momentum must change is $\dot{L}_{orb} - \dot{L}_{orb,T}$, i.e., $(\gamma/2)a'^{1/2}(\dot{\varepsilon}_T - \dot{\varepsilon}_s)/(1 - \varepsilon_s)^{1/2}$. These agree if eqn. (5.8) is used to evaluate $\sin 2\theta_s = (\dot{\varepsilon}_T - \dot{\varepsilon}_s)/20\chi\alpha\varepsilon_s$. However, since the extreme value of $\sin 2\theta_s \approx (\dot{\varepsilon}_T - \dot{\varepsilon}_*)/20\chi\alpha\varepsilon_* \to -1$, the strongest possible torque is $T'_{max} = -10\chi\alpha\gamma a'^{1/2}\varepsilon_*$. Accordingly, the resonance could not be maintained if

$$\chi \equiv \Omega_\odot t_T < -(\dot{\varepsilon}_T - \dot{\varepsilon}_s)/20\alpha\varepsilon_*(1 - \varepsilon_*) \equiv \chi_{crit} \tag{B5}$$

**Appendix C - Mignard Tidal Model**

Mignard first derives the force due to a second-order tidal distortion raised on the Earth by the Moon in the vector form,

$$F = -3k_T \frac{Gm^2 R^5}{r^{10}} \Delta t[2(\mathbf{r} \cdot \mathbf{v})\mathbf{r} + r^2(\mathbf{r} \times \mathbf{s} + \mathbf{v})] \tag{C1}$$

Where $k_T$ is the tidal Love number for the Earth, vectors $\mathbf{r}, \mathbf{v}$ are the position and velocity of the Moon of mass $m$, and $\mathbf{s}$ is the Earth's spin vector, which, for simplicity, we will assume is perpendicular to the lunar orbit plane. The radial, $F_r$, and tangential, $F_\theta$, force components are then substituted into Gauss' form of the Lagrange equations (*e.g.,* Brouwer and Clemence, 1961),

$$\frac{da}{dt} = \frac{2}{mn(1-e^2)^{1/2}} \left[ F_r e \sin\theta + F_\theta \frac{p}{r} \right]; \quad \frac{de}{dt} = \frac{(1-e^2)^{1/2}}{man} \left[ F_r \sin\theta + F_\theta \frac{1}{e}\left(\frac{p}{r} - \frac{r}{a}\right) \right] \quad \text{(C2a,b)}$$

where $p \equiv a(1 - e^2)$ and the rates are then averaged over an orbit to give the tidal changes in semi-major-axis and eccentricity.

**Appendix D - Permanent Figure Torque**

Consider a Moon with principal moments of inertia $C_m \geq B_m \geq A_m$, where $C_m$ is the moment about its spin axis, assumed to be normal to its orbit plane, and $A_m$ is the moment about the Moon's long axis. The instantaneous value of the permanent figure (*pf*) torque is given by Danby (1992; see also Murray and Dermott, 1999),

$$T_{pf} = -\frac{3}{2}(B_m - A_m)(GM/r^3) \sin 2\psi = C_m \frac{ds_{m,pf}}{dt} \quad \text{(D1)}$$

where $r$ is the Earth-Moon distance, $\psi$ is the angle between the long axis of the Moon and the Earth-Moon line, *i.e*, $\psi = \vartheta - f$, where $\vartheta$ is the angular position of the Moon's long axis with respect to the perigee, $\varpi$, and $f$ is the true anomaly (*e.g.*, Goldreich and Peale, 1966a,b). We set $\vartheta = s_m t + \psi_o$, which for synchronous rotation is $\vartheta = nt + \psi_o$, where $\psi_o$ is the value of $\psi$ at perigee. If $T_{pf}$ is then averaged over an orbit, one obtains (*e.g.*, Goldreich and Peale, 1966a,b),

$$\langle T_{pf} \rangle = -\frac{3}{2} n^2 (B_m - A_m) H(\varepsilon) \sin 2\psi_o \quad \text{(D2)}$$

where $H(\varepsilon) = 1 - 5\varepsilon/2 + 13\varepsilon^2/16$ is a so-called Hansen polynomial. This torque leads to further contributions to semi-major axis and eccentricity variations, $\dot{a}'_{pf}$ and $\dot{\varepsilon}_{pf}$.

Analogous to eqn. (4.2), conservation of angular momentum requires

$$\dot{s}'_{m,pf} = -\frac{\gamma}{2\kappa} a'^{1/2} (1 - \varepsilon)^{1/2} \left( \frac{\dot{a}'_{pf}}{a'} - \frac{\dot{\varepsilon}_{pf}}{1-\varepsilon} \right). \quad \text{(D3)}$$

This must (nearly) balance the tidal torque to ensure synchronous stability so that $\dot{s}'_m + \dot{s}'_{m,pf} = \dot{n}/\Omega_\oplus$, and the off-set angle $\psi_o$ adopts the value needed to accomplish this.

A major difference between a torque on the permanent figure of the Moon and a torque on a tidal distortion is that the former is not accompanied by energy dissipation due to planetary flexing. Accordingly, the combination of orbital energy and spin energy of the Moon is also conserved under its action, *i.e., $d(\kappa\lambda s'^2_m/2 - \mu/2a')/d\tau|_{PF} = 0$*. Taking the derivatives and rearranging yields an additional condition,

$$\dot{s}'_{m,pf} = -\frac{\gamma}{2\kappa} \frac{\dot{a}'_{pf}}{s'_m a'^2} = -\frac{\gamma}{2\kappa} a'^{1/2} \frac{\dot{a}'_{pf}}{a'} \quad \text{(D4)}$$

where the final expression sets $s'_m = n/\Omega_\oplus = a'^{-3/2}$.

Approximating $\dot{s}'_m + \dot{s}'_{m,pf} \approx 0$ due to the smallness of $\dot{n}$ compared to either spin acceleration, we conclude that

$$\left(\frac{\dot{a}'_{pf}}{a'} - \frac{\dot{\varepsilon}_{pf}}{1-\varepsilon}\right) \approx -\left(\frac{\dot{a}'_m}{a'} - \frac{\dot{\varepsilon}_m}{1-\varepsilon}\right) \tag{D5}$$

However, one cannot simply assume equal but opposite values for $\dot{a}'_{pf} = -\dot{a}'_m$ and $\dot{\varepsilon}_{pf} = -\dot{\varepsilon}_m$, because the permanent figure torque may partition its changes in $a$ and $\varepsilon$ differently than do tides. From eqns. (D3) and (D4) we get

$$\frac{[1-(1-\varepsilon)^{1/2}]}{(1-\varepsilon)^{1/2}} \frac{\dot{a}'_{pf}}{a'} = -\frac{\dot{\varepsilon}_{pf}}{1-\varepsilon} \tag{D6}$$

Using this to eliminate either $\dot{a}'_{pf}$ or $\dot{\varepsilon}_{pf}$ in eqn. (D5) leads to,

$$\frac{\dot{a}'_{pf}}{a'} = -f_{pf}\left(\frac{\dot{a}'_m}{a'} - \frac{\dot{\varepsilon}_m}{1-\varepsilon}\right) \;;\; \dot{\varepsilon}_{pf} = g_{pf}\left(\frac{\dot{a}'_m}{a'} - \frac{\dot{\varepsilon}_m}{1-\varepsilon}\right) \tag{D7a,b}$$

where $f_{pf} \equiv (1-\varepsilon)^{1/2}$ and $g_{pf} = (1-\varepsilon)[1-(1-\varepsilon)^{1/2}]$. The total change rates for $a'$ and $\varepsilon$ due to both tides and $T_{pf}$ for a Moon in synchronous rotation is then

$$\frac{\dot{a}'}{a'} = \frac{\dot{a}'_\oplus}{a'} + (1-f_{pf})\frac{\dot{a}'_m}{a'} + f_{pf}\frac{\dot{\varepsilon}_m}{1-\varepsilon} \tag{D8}$$

$$\dot{\varepsilon}_T = \dot{\varepsilon}_\oplus + \left(1 - \frac{g_{pf}}{1-\varepsilon}\right)\dot{\varepsilon}_m + g_{pf}\frac{\dot{a}'_m}{a'}, \tag{D9}$$

where $s'_m a'^{3/2} = 1$, which in combination with (4.9a,b) gives

$$\dot{a}'_m/a' = A[f_1(\varepsilon) - f_2(\varepsilon)]/a'^8 \;;\; \dot{\varepsilon}_m = A\varepsilon[g_1(\varepsilon) - g_2(\varepsilon)]/a'^8 \tag{D10a,b}$$

The above rates are valid so long as synchronous rotation can be maintained. However, $|\sin 2\psi_o|$ has a maximum value of unity, and so from eqns. (4.2) and (D2) there is a minimum value required for $(B_m - A_m)/C_m$,

$$\frac{(B_m - A_m)}{C_m} > \left|\frac{\gamma}{3\kappa}\left(\frac{a'^{7/2}}{\Omega_\oplus \tau_T}\right)\frac{(1-\varepsilon)^{1/2}}{H(\varepsilon)}\left(\frac{\dot{a}'_m}{a'} - \frac{\dot{\varepsilon}_m}{1-\varepsilon}\right)\right|$$

$$= \left|\frac{\gamma}{3\kappa}\left(\frac{A}{\Omega_\oplus \tau_T}\right)\frac{1}{a'^{9/2}}\frac{(1-\varepsilon)^{1/2}}{H(\varepsilon)}\left[f_1 - f_2 - \frac{\varepsilon}{1-\varepsilon}(g_1 - g_2)\right]\right| \tag{D11a}$$

where the final expression sets $s'_m a'^{3/2} = 1$. This criterion reads

$$\frac{(B_m - A_m)}{C_m} > 4 \times 10^{-4}\left(\frac{k_m \Delta t_m}{4\text{ min}}\right)\frac{(1-\varepsilon)^{1/2}}{H(\varepsilon)}\left[f_1 - f_2 - \frac{\varepsilon}{1-\varepsilon}(g_1 - g_2)\right]\left(\frac{7}{a'}\right)^{9/2} \tag{D11b}$$

where $k_m \Delta t_m \approx 4$ min for the current Moon (Williams and Boggs, 2015). If violated, the synchronous lock is broken.

The above estimate considers whether the permanent figure torque is sufficient to maintain synchronous rotation against the competing tidal torque. Goldreich (1966) considered an initial rotation faster than $n$, and found that this rate would decrease, librate about synchronous rotation, and ultimately damp to the synchronous state if

$$\frac{(B_m - A_m)}{C_m} \gtrsim 7.5\pi^2 \varepsilon^2 \qquad (D12)$$

For a shape similar to that of the current Moon, with $(B_m - A_m)/C_m = 2.28 \times 10^{-4}$, eqn. (D11b) implies that synchronous lock could be maintained at the time the Moon encounters evection (*i.e.*, $a' \sim 7$) for an initially low eccentricity ($\varepsilon < 0.095$), but that non-synchronous rotation would ensue as $e$ became large. Eqn. (D12) implies that the $(B_m - A_m)/C_m$ of the current Moon would be sufficient to establish synchronous rotation for $\varepsilon < 0.0018$. Of course, the current $(B_m - A_m)/C_m$ value may not have pertained to the early Moon, and so it is prudent to consider both synchronous and non-synchronous cases.

For the case of a non-synchronously rotating Moon without permanent figure torques, eqns. (6.10) and (6.12) in the main text provide the partial derivatives needed to evaluate whether the libration amplitude grows or damps. Analogous expressions can be developed for synchronous lunar rotation maintained by a permanent figure torque, with $\partial(\dot{\varepsilon}_T)/\partial\varepsilon$ replaced by $\partial(\dot{\varepsilon}_T + \dot{\varepsilon}_{pf})/\partial\varepsilon$, and $\partial(\dot{a}'/a')/\partial\varepsilon$ replaced by $\partial(\{\dot{a}' + \dot{a}'_{pf}\}/a')/\partial\varepsilon$, with $\dot{\varepsilon}_{pf}$ and $\dot{a}'_{pf}$ given in (D7). These are

$$\frac{a'^8}{A} \frac{\partial \dot{\varepsilon}_{pf}}{\partial \varepsilon} = (f_1 - f_2)\frac{\partial g_{pf}}{\partial \varepsilon} + \left(\frac{\partial f_1}{\partial \varepsilon} - \frac{\partial f_2}{\partial \varepsilon}\right) g_{pf}$$

$$-(g_1 - g_2)\left(\frac{\partial g_{pf}}{\partial \varepsilon}\frac{\varepsilon}{1-\varepsilon} + \frac{\varepsilon g_{pf}}{(1-\varepsilon)^2} + \frac{g_{pf}}{1-\varepsilon}\right) - \left(\frac{\partial g_1}{\partial \varepsilon} - \frac{\partial g_2}{\partial \varepsilon}\right)\frac{\varepsilon g_{pf}}{1-\varepsilon} \qquad (D13)$$

and

$$\frac{a'^8}{A} \frac{\partial}{\partial \varepsilon}\left(\frac{\dot{a}'_{pf}}{a'}\right) = \frac{1}{(1-\varepsilon)^{\frac{1}{2}}} \qquad (D14)$$

$$\left[\frac{f_1 - f_2}{2} + (g_1 - g_2)\left(1 + \frac{\varepsilon}{2(1-\varepsilon)}\right) + \varepsilon\left(\frac{\partial g_1}{\partial \varepsilon} - \frac{\partial g_2}{\partial \varepsilon}\right) + \left(\frac{\partial f_2}{\partial \varepsilon} - \frac{\partial f_1}{\partial \varepsilon}\right)(1 - \varepsilon)\right],$$

where $\frac{\partial g_{pf}}{\partial \varepsilon} = \frac{3}{2}(1 - \varepsilon)^{1/2} - 1$.

## Appendix E - Additional Zero Libration Evolutions

Here we show additional zero-libration evolutions as considered in Section 5. Note that in these and the other evolutions in the main text we ignore the potential for tidal disruption when the lunar perigee is interior to the Roche limit, which can occur for low $A$ cases.

Figure A1 displays tracks for $A = 10$ with different starting values of the system angular momentum, $L'_0$, corresponding to varied initial Earth spin rates, $s'_0$, following a lunar forming impact. Changing $L'_0$ alters the encounter distance for the resonance as in Figure 4. For lower $L'_0$, the resonance occurs closer to the Earth and the stall in the Moon's orbital expansion occurs at smaller $a'_c$ and $\varepsilon_c$. However, all cases eventually converge on the same end state in the limiting case that the Moon remains in resonance throughout its whole evolution (which as we show in Section 6 is unlikely to occur, as much earlier resonance escape is predicted). Accordingly, the higher the starting $L'_0$, the greater the angular momentum decay, $\Delta L' = L'_0 - L'_f$, and evolutionary tracks for high $L'_0$ are reminiscent of those shown in CS12.

Figure A2 compares evolutionary tracks with $L_0 = 2L_{EM}$ for other values of $A$. As $A$ increases, the stationary state eccentricity is suppressed by progressively stronger lunar tides. This in turn weakens the tidal torque (due to the larger lunar periapsis), prolonging the evolutionary time scale. Figure A3 displays a synchronous evolution with $A = 10$, $L_0 = 2\,L_{EM}$ contrasted to the non-synchronous evolution shown in Figure 4 in the main text, shown in grey. Here we have set $s'_m a'^{3/2} = 1$, and modified the expressions for tidal changes in $a$ and $\varepsilon$ to include the permanent figure torques as in eqns. (D8) and (D9). The non-synchronous track acquires higher maximum values for $a$ and $\varepsilon$ but these then decrease somewhat more rapidly than in the synchronous case.

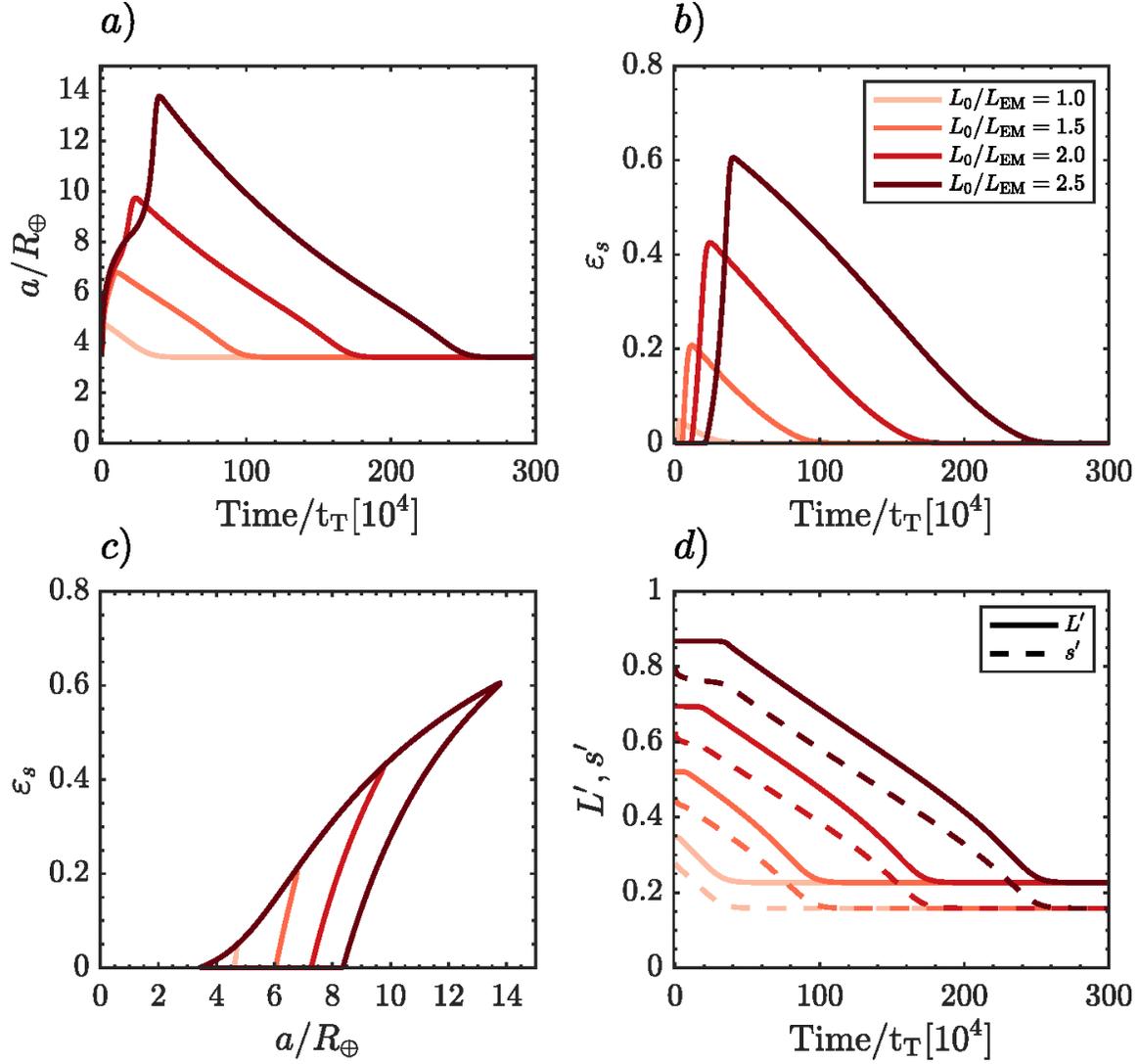

**Figure A1.** System evolution with $A=10$ for various values of $L_o$, assuming a Moon in non-synchronous rotation.

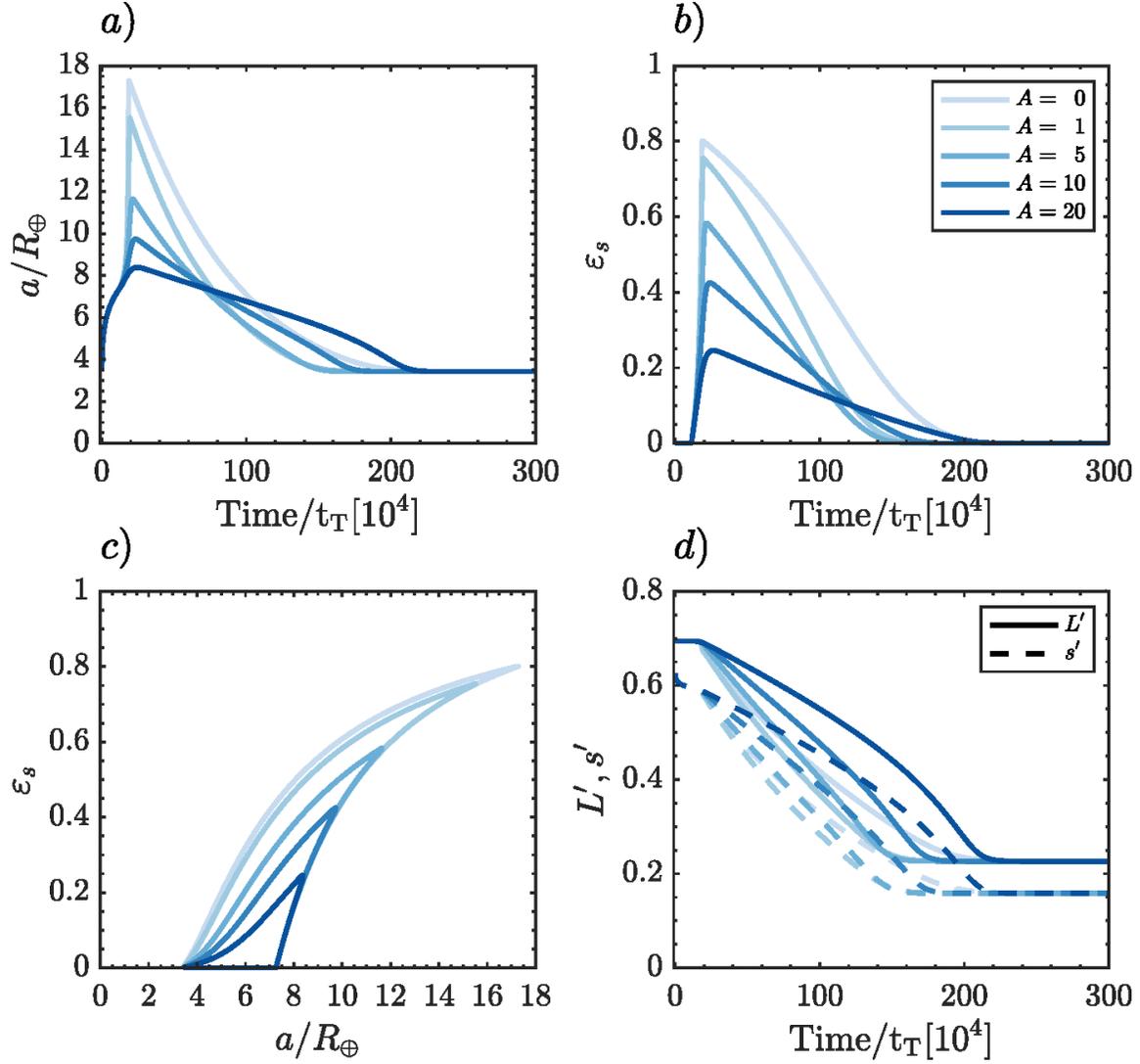

**Fig. A2.** System evolution for various values of $A$ with $L_o = 2L_{EM}$, assuming a Moon in non-synchronous rotation.

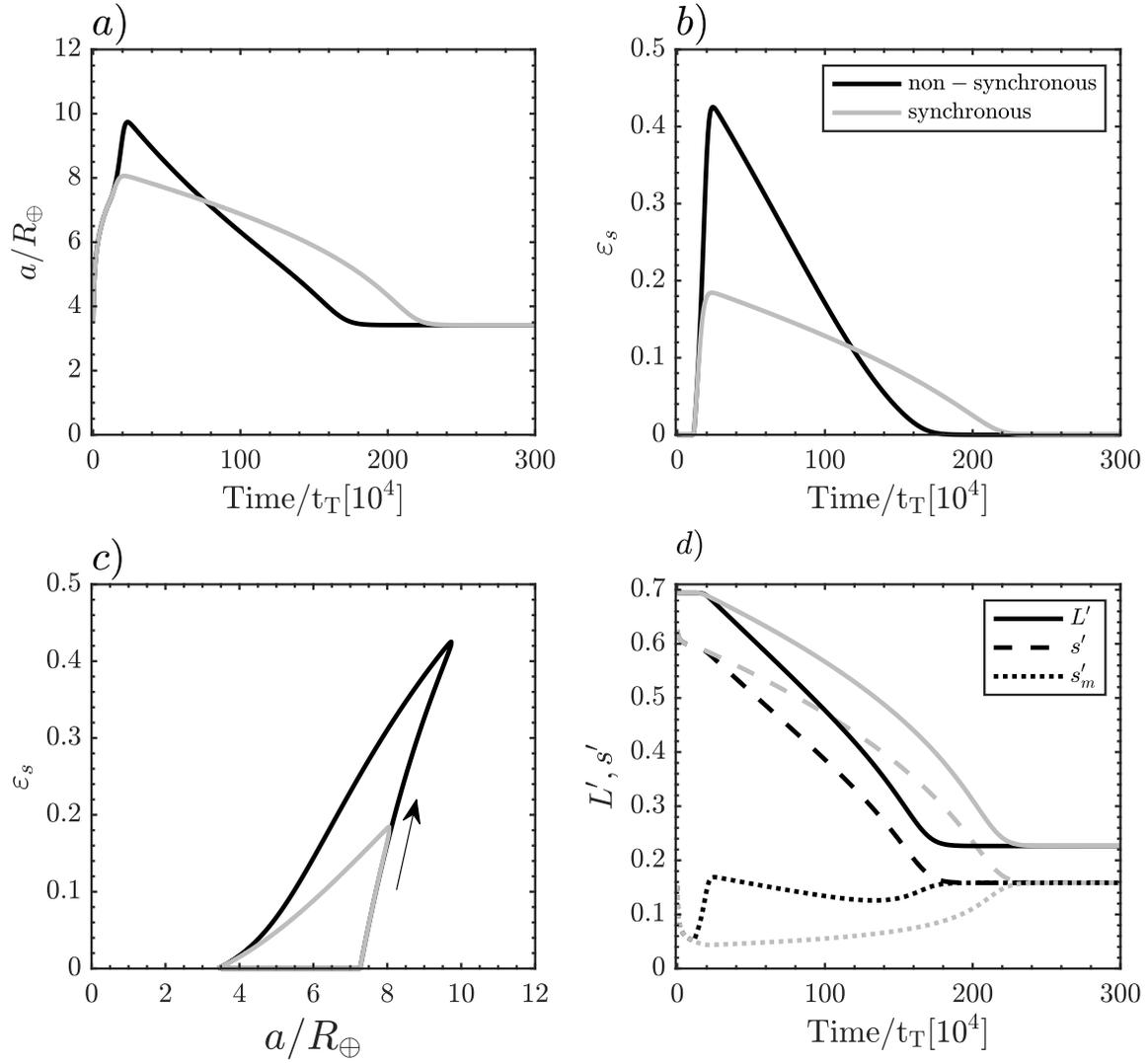

**Fig. A3.** System evolution for a Moon with synchronous rotation maintained by a permanent figure torque with $L_o = 2L_{EM}$ and $A = 10$ (grey), with non-synchronous rotation case from Figure 4 in the main text shown for comparison (black).